\documentclass[]{emulateapj}
\usepackage{epsfig, natbib, graphicx, color}

\input epsf

\newcommand{\Mpc}{\mbox{Mpc}}

\newcommand{\bm}[1]{\mathbf{#1}}

\newcommand{\avgn}{\langle \bar n \rangle}

\newcommand{\na}{\bar n_a}
\newcommand{\nap}{\bar n_{a'}}

\newcommand{\psia}{\psi_a}
\newcommand{\psiap}{\psi_{a'}}

\newcommand{\avgpsia}{\langle \psia|m \rangle}

\newcommand{\Na}{N_a}
\newcommand{\Nap}{N_{a'}}

\newcommand{\nm}{ \frac{d\avgn}{dm}}

\newcommand{\avgna}{\langle \na \rangle}

\newcommand{\Nt}{N_{t}}

\newcommand{\Nobs}{N_{obs}}

\newcommand{\Nsat}{N_{sat}}

\newcommand{\avg}[1]{\left\langle #1 \right\rangle}

\newcommand{\pnn}{P(\Nobs|\Nt)}
\newcommand{\ps}{P_s(\Nobs|\Nt)}
\newcommand{\pn}{P_n(\Nobs|\Nt)}
\newcommand{\corr}{C_{a,a'}}
\newcommand{\lk}{{\cal{L}}}
\newcommand{\zh}{z_h}

\newcommand{\zc}{z_c}

\newcommand{\zmin}{z_{min}}
\newcommand{\zmax}{z_{max}}

\newcommand{\Ns}{N_s}
\newcommand{\bn}{\hat \bm{n}}
\newcommand{\LCDM}{\Lambda\mbox{CDM}}

\shortauthors{ROZO ET AL.}
\shorttitle{OPTICALLY-SELECTED CLUSTERS AS COSMOLOGICAL TOOL}

\begin{document}
\title{Optically-Selected Cluster Catalogs as a Precision Cosmology Tool}
\author{Eduardo Rozo \altaffilmark{1,2,3}, 
Risa H. Wechsler\altaffilmark{3,4}, 
Benjamin P. Koester\altaffilmark{5,6}, 
August E. Evrard\altaffilmark{5}, 
Timothy A. McKay\altaffilmark{5}
}

\altaffiltext{1}{CCAPP, The Ohio State University, Columbus, OH 43210, erozo@mps.ohio-state.edu}
\altaffiltext{2}{Department of Physics, The University of Chicago, Chicago, IL 60637}
\altaffiltext{3}{
  Kavli Institute for Cosmological Physics,
  The University of Chicago, 
  Chicago, IL 60637
}
\altaffiltext{4}{
  Kavli Institute for Particle Astrophysics \& Cosmology,
  Physics Department, and Stanford Linear Accelerator Center,
  Stanford University,
  Stanford, CA 94305
}
\altaffiltext{5}{Physics Department, University of Michigan, Ann
  Arbor, MI 48109}
\altaffiltext{6}{Department of Astronomy, The University of Chicago, Chicago, IL 60637}

\begin{abstract}
We introduce a framework for describing the halo
selection function of optical cluster finders.  We treat the problem as
being separable into a term that describes the intrinsic galaxy content of a 
halo (the Halo Occupation Distribution, or HOD) and a term that captures the effects 
of projection and selection by the particular cluster finding algorithm. 
Using mock galaxy catalogs tuned to reproduce the luminosity dependent correlation
function and the empirical color-density relation measured in the SDSS,
we characterize the maxBCG algorithm applied by Koester et al. to the SDSS galaxy catalog. 
We define and calibrate measures of completeness and purity for this algorithm, and 
demonstrate successful recovery of the underlying cosmology and HOD when applied to 
the mock catalogs.   We identify principal components --- combinations of cosmology and 
HOD parameters --- that are recovered by survey counts as a function of richness,  and 
demonstrate that percent-level accuracies are possible in the first two components, if 
the selection function can be understood to $\sim 15\% $ accuracy. 
\end{abstract}

\section{Introduction}

It has long been known that the abundance of massive halos in the
universe is a powerful cosmological probe.  From theoretical
considerations \citep[][]{pressschechter74,bondetal91,shethtormen02}
one expects the number of massive clusters in the universe to be
exponentially sensitive to the amplitude of the matter power spectrum
$\sigma_8$, a picture that has been confirmed with extensive numerical
simulations
\citep[e.g.][]{jenkinsetal01,shethtormen02,warrenetal05}.\footnote{Here,
  we characterize the present day amplitude of the power spectrum with
  the usual parameter $\sigma_8$, the rms amplitude of density
  perturbations in spheres of $8h^{-1}\ \Mpc$ radii.}  Moreover, since
the number of halos also depends on the mean matter density of the
universe, cluster abundance constraints typically result in
degeneracies of the form $\sigma_8\Omega_m^\gamma \approx constant$
where $\gamma\approx 0.5$.  This type of constraint is usually
referred to as a cluster normalization condition\citep[see e.g.][for a
discussion of the origin of this degeneracy]{rozoetal04}.

There are, however, important difficulties one must face in
determining $\sigma_8$ from any given cluster sample.  Specifically,
the fact that cluster masses cannot be directly observed implies that
some other observable such as X-ray emission or galaxy overdensity
must be relied upon both to detect halos and estimate their
masses. Consequently, characterizing how what one sees, the cluster
population, is related to what we can predict, the halo population, is
of fundamental importance.  In fact, it is precisely these types of
systematic uncertainties that dominate the error budget in current
cosmological constraints from cluster abundances \citep[see
e.g.][]{seljak02,pierpaolietal03,henry04}.

Optical surveys are traditionally thought of as being particularly
susceptible to these types of systematics, a belief that is largely
historical in origin.  The earliest cluster catalogs available were
created through visual identification of galaxy clusters
\citep[][]{abell58,zwickyetal68, shectman85,aco89,gunnetal86}, and
thus cluster selection was inherently not quantifiable.  While the situation
was much improved by the introduction of automated cluster-finding
algorithms
\citep[][]{shectman85,lumsdenetal92,daltonetal97,galetal00,galetal03},
projection effects --- the identification of spurious concentrations
of galaxies along the line of sight as physical groupings --- remained
a significant obstacle \citep[see
e.g.][]{lucey83,katgertetal96,postmanetal96,vanhaarlemetal97,okeetal98}.
These projection effects can be minimized by turning to spectroscopic
surveys, though even then difficulties arise due to the finger-of-god
elongation along the line of sight
\citep[][]{huchrageller82,noltheniuswhite87,mooretal93,ramellaetal89,
  kochaneketal03,merchanzandivarez02,ekeetal04a,
  yangetal05a,milleretal05,berlindetal06}.  Alternatively, several new
optical cluster-finding algorithms have been developed that take
advantage of the accurate photometry available in large digital sky
surveys such as the Sloan Digital Sky Survey \citep[SDSS,
][]{yorketal00} to largely, though not completely, overcome this
difficulty \citep[][]{kepneretal99,gladdersyee00,whitekochanek02,gotoetal02,kimetal02,
  koesteretal06a}.

The challenge that confronts optical cluster work today is to
demonstrate that these type of selection effects can be, if not
entirely overcome, then at least properly taken into account within
the context of parameter estimation in cosmological studies.  In this
work, we introduce such a scheme.  The key idea behind our analysis is
to define the cluster selection function $P(\Nobs|m)$ as the
probability that a halo of mass $m$ be detected as cluster with
$\Nobs$ galaxies, and then make the fundamental assumption that
cluster detection is a two-step process: first, there is a probability
$P(\Nt|m)$ that a mass $m$ halo will contain $\Nt$ galaxies, and
second, there is a matrix $P(\Nobs|\Nt)$ which describes the
probability that a halo with $\Nt$ galaxies will be detected as a
cluster with $\Nobs$ galaxies.  In other words, we are assuming that
cluster detection depends on mass primarily through the number of galaxies 
hosted by the cluster's halo.  Note that the probability $P(\Nobs|\Nt)$
characterizes not only measurement errors but any possible systematic
errors such as line of sight projections. A key advantage of defining the
problem in this way is that it leads one naturally to precise
definitions of purity and completeness for a given cluster sample, and
allows for proper marginalization of our results over all major
systematic uncertainties.

The formalism outlined here could be generally applied to any
optically-identified cluster survey, but we focus herein on its
application to the SDSS maxBCG catalog.  In particular, we test the
method by populating dark matter simulations with galaxies as
described in \citet[][]{wechsleretal07} and then running the maxBCG
cluster finding algorithm of \citet[][]{koesteretal06a} in the
resulting mock galaxy catalog.  Note that the resulting cluster
catalog will suffer all of the major systematics affecting the
corresponding data catalog from \citet[][]{koesteretal06b}, including
incompleteness (non detections), impurities (false detections), and
systematic biasing of galaxy membership in clusters from galaxies
projected along the line of sight.  By comparing the underlying halo
population to the resulting cluster catalog we characterize the maxBCG
cluster selection function in the mock catalogs.  Using a maximum
likelihood analysis, we then demonstrate that when the cluster
selection function is known at a quantitative level, we can
successfully recover the cosmological and HOD parameters of each of
the mocks to within the intrinsic degeneracies of the data.  We
emphasize that these results explicitly demonstrate that our analysis
correctly takes into account the systematic uncertainties inherent to
the data.

The layout of the paper is as follows. In \S \ref{sec:model} we
describe our model, including our parameterization of the various
systematic uncertainties that affect real cluster samples.
The quantitative calibration of the
cluster selection function for the maxBCG cluster finding algorithm of
\citet[][]{koesteretal06a} through the use of numerical simulations
is detailed in \S \ref{sec:calibration}.
In \S \ref{sec:model_testing} we investigate whether our model
accurately describes the cluster selection function, and in
particular, whether we can recover the cosmological parameters of mock
cluster samples using the techniques developed in this paper.  
We summarize in \S \ref{sec:conclusions}.


\section{The Model}
\label{sec:model}

This section describes our general framework in detail. We begin by
considering a perfect cluster finding algorithm, and slowly add the
various layers of complexity that arise in the real world.  We first
allow observational scatter in cluster richnesses, and
demonstrate that this naturally gives rise to the concepts of purity
and completeness.  We then include the effects of photometric redshift
uncertainties, finally discussing how a careful calibration of these
various difficulties can be included within a maximum likelihood
analysis of an observational data set.

\subsection{The Basic Picture}

The basic tenet of our model is that galaxy clusters are associated
with massive halos. Consider then a cluster sample where $\Nobs$ is
used to denote the number of observed galaxies within the cluster.  We
refer to $\Nobs$ as the cluster's \it richness. \rm If $P(\Nobs|m)$ is
the probability that a mass $m$ halo has $\Nobs$ galaxies, and there
are $(d\avg{\bar n}/dm)dm$ such halos, the number of clusters with
$\Nobs$ galaxies is simply
\begin{equation}
\langle\bar{n}(\Nobs)\rangle = \int dm\ \nm P(\Nobs|m)
\end{equation}

If one were to bin the data such that bin $a=[R_{min},R_{max})$ contains
all clusters with $R_{max}>\Nobs\geq R_{min}$, one need only sum the above
expression over the relevant values of $\Nobs$.  As we shall see momentarily,
it is useful to define a binning function $\psia(\Nobs)$ such that 
$\psia(\Nobs)=1$ if $\Nobs$ falls in bin $a$ and zero otherwise.  With this
definition, the number of clusters in bin $a$ can be re-expressed as
\begin{equation}
\avgna = \int dm\ \frac{d\avgn}{dm}\avgpsia.
\label{eq:clden}
\end{equation}
where $\avgpsia$ contains the sum over all $\Nobs$ and is defined as
\begin{equation}
\avgpsia = \sum_{\Nobs}P(\Nobs|m)\psia(\Nobs).
\label{eq:old_binning}
\end{equation}

Proper modeling of an observational sample reduces to understanding the
probability distribution $P(\Nobs|m)$.



We consider first the case of a perfect cluster-finding algorithm:
assume the algorithm detects all halos, there are no false detections,
and that the observed number of galaxies $\Nobs$ is equal to the true
number of halo galaxies $\Nt$ for every halo.  The probability
$P(\Nt|m)$ is called the Halo Occupation Distribution (HOD), and
characterizes the intrinsic scatter in the richness-mass relation.
Following \citet[][see also
\citealt{zhengetal05,yangetal05b}]{kravtsovetal04} we assume that the
total number of galaxies in a halo takes the form $\Nt=1+\Nsat$ where
$\Nsat$, the number of satellite galaxies in the cluster, is Poisson
distributed at each $m$ with an expectation value $\avg{\Nt|m}$ given
by
\begin{equation}
\avg{\Nsat|m}= \left(\frac{m}{M_1}\right)^\alpha.
\label{eq:hod}
\end{equation}
Here, $M_1$ is the characteristic mass at which halos acquire one
satellite galaxy.  Note that in cluster abundance studies, the
typical mass scale probed is considerably larger than $M_1$.  Nevertheless,
the above parametrization is convenient since degeneracies between
HOD and cosmological parameters take on particularly simple forms when 
parameterized in this way \citep[see][]{rozoetal04}.


\subsection{Noise in Galaxy Membership Assignments}

In general, the observationally-determined number of galaxies
$\Nobs$ in a cluster may differ from the true number of galaxies $\Nt$
in the corresponding halo.  That is to say, we expect there is a
probability distribution $P(\Nobs|\Nt)$ that gives us the probability
that a halo with $\Nt$ galaxies will be detected as a cluster with
$\Nobs$ galaxies.  Before we look in more detail at the probability
matrix $P(\Nobs|\Nt)$, we investigate how the above assumption affects
the final expression for cluster abundances. Equations \ref{eq:clden}
and \ref{eq:old_binning} are, of course, unchanged, though the
probability matrix $P(\Nobs|m)$ is no longer identical to the halo
occupation distribution $P(\Nt|m)$.  Rather, it is related to the HOD
via $P(\Nobs|m) = \sum_{\Nt} P(\Nobs|\Nt)P(\Nt|m)$.  Consequently, the
quantity $\avg{\psia|m}$ becomes
\begin{equation}
\avg{\psia|m}  = \sum_{\Nt}\tilde\psia(\Nt) P(\Nt|m)
\label{eq:new_binning1}
\end{equation}
where $\tilde\psia(\Nt)$ is defined as
\begin{equation}
\tilde\psia(\Nt) = \sum_{\Nobs} \psia(\Nobs)P(\Nobs|\Nt).
\label{new_binning2}
\end{equation}

Since equation \ref{eq:new_binning1} has the same form as equation 
\ref{eq:old_binning}, as long as $P(\Nobs|\Nt)$ is known
we can view observational errors simply as a re-binning of the data.


\subsection{Completeness}
\label{sec:completeness}

In general, we expect non-zero matrix elements in the matrix $P(\Nobs|\Nt)$
to arise in one of two ways:

\begin{enumerate}
\item The cluster-finding algorithm worked correctly: it detected a cluster 
where there is
a halo, and the assigned richness $\Nobs$ is close to its expected value 
$\avg{\Nobs|\Nt}$.  
\item The cluster-finding algorithm worked incorrectly: it either failed to detect a
cluster where there was a halo ($\Nobs =0,\ \Nt\neq 0$), detected a cluster where
there were no halos ($\Nobs\neq 0,\ \Nt=0$), or the richness estimate was grossly
incorrect.
\end{enumerate}

Imagine marking now every non-zero matrix element of the matrix
$P(\Nobs|\Nt)$ with a point on the $\Nobs-\Nt$ plane (see Figure
\ref{fig:pnn} for an example).  In general, we expect points for which
the cluster-finding algorithm worked correctly to populate a band
around the expectation value $\avg{\Nobs|\Nt}$, which we refer to as
the \it signal band. \rm Points falling outside the signal band we
refer to as noise, and represent those instances where the
cluster-finding algorithm suffered a catastrophic error.  Generically,
we expect that the values of the probability matrix $\pnn$ within the
signal band will be stable and easy to characterize, whereas the noise
part of the matrix will be unstable and difficult to characterize.
Our challenge is then to come up with a reasonable way to account for
the noise part of the probability matrix in cluster abundance studies.

Let us begin our attack on this problem with some definitions.  We
define the quantity $c(\Nt)$ as the probability that a halo with $\Nt$
galaxies be correctly detected.  Thus, $c(\Nt)$ is simply the sum of
all matrix elements $P(\Nobs|\Nt)$ within the signal band at fixed
$\Nt$.  Note since $c(\Nt)$ is the probability of a halo being
detected as signal, the expectation value for the fraction of signal
halos is precisely $c(\Nt)$.  We thus refer to $c(\Nt)$ as the \it
completeness function. \rm We emphasize, however, that $c(\Nt)$ is
fundamentally a probability, and consequently it contributes to the
correlation matrix of the observed cluster counts.  We also define the
\it signal matrix \rm $\ps$ via
\begin{equation}
\ps = \pnn/c(\Nt)
\label{eq:sigmat}
\end{equation}
for matrix elements within the signal band, and $\ps=0$ otherwise.  In other words,
the signal matrix is what the probability matrix would be if there was no 
noise (catastrophic errors) 
in the data.  Finally, we define the noise matrix $\pn$ to be zero within
the signal band, and equal to $\pnn$ otherwise.  We can thus write
\begin{equation}
\pnn = c(\Nt)\ps+\pn.
\end{equation}

Inserting this expression into equations
\ref{eq:clden} and \ref{eq:new_binning1}, we see that the total abundance
is a sum of a signal term and a noise term, 
\begin{equation}
\avg{\na}=\avg{\na}_s + \avg{\na}_n.
\end{equation}
If one is willing to drop the information contained within the noise term,
then characterizing $\pnn$ in the noise regime become unnecessary.  All one
needs to do instead is characterize the completeness $c(\Nt)$ and the
noise contribution $\avg{\na}_n$.  

\subsection{Purity}
\label{sec:purity}

We take a probabilistic approach for characterizing the noise
contribution $\avg{\na}_n$ to the cluster density.  Specifically, we
define the \it purity function \rm $p(\Nobs)$ as the probability that
a cluster with $\Nobs$ galaxies be signal.  Consider then a fixed
richness $\Nobs$, and let $N$ be the number of observed clusters and
$N_s$ be the number of signal clusters.  The expectation value for
$N_s$ given $N$ is thus $\avg{N_s|N}=Np$ where $p$ is the purity.
Note, however, that this is \it not \rm the quantity we are interested
in.  For modeling purposes, \it we are interested in the number of
observed clusters $N$ given the predicted number of signal clusters
$N_s$. \rm That is, we need to compute $\avg{N|N_s}$. To compute this
number, we first note that the probability distribution $P(N_s|N)$ is
a simple binomial distribution
\begin{equation}
P(N_s|N) = \left(\begin{array}{c} N \\ N_s \end{array} \right)
	p^{N_s}(1-p)^{N-N_s}.
\end{equation}
In the limit $N\gg 1$, we can approximate the binomial distribution as a 
Gaussian with expectation value $\avg{\Ns|N}=Np$ and variance 
$\mbox{Var}(\Ns|N)=Np(1-p)$.  By Bayes's theorem, $P(N|\Ns)$ is simply
proportional to $P(\Ns|N)$, so we find
\begin{equation}
P(N|\Ns) = \frac{A}{\sqrt{2\pi Np(1-p)}}\exp\left(-\frac{(N-\Ns/p)^2}{2N(1-p)/p}\right)
\end{equation}
where $A$ is a normalization constant.  We can further simplify this
expression in the limit $\Ns\gg 1$ and for cases where the purity is
close to unity.  In this limit, the probability distribution
$p(\Ns|N)$ becomes very narrow, and the expectation value for the
random variable $x$ defined via $\Ns (1+x)=N$ will be close to zero.
Expanding the above expression around $x=0$ and keeping only the
leading order terms we obtain
\begin{equation}
\rho(x|\Ns) = A\exp\left(-\frac{(x-\mu)^2}{2\sigma^2}\right)
\end{equation}
where $\mu=(1-p)/p$ and $\sigma^2=\mu/\Ns$.  Note that since
$\sigma^2\ll \mu$, we can extend the range of $x$ to vary from
$-\infty$ to $+\infty$, in which case $\rho(x|\Ns)$ becomes a simple
Gaussian.  It follows that the expectation value $\avg{N|\Ns}$ is
given by $\avg{N|\Ns}=\Ns/p$, exactly what we would expect.

At this point it might seem that the above argument was really an
unnecessary complication in that our end result is exactly what we
would have naively guessed.  Nevertheless, our argument is useful in
that it proves that this guess is indeed correct.  Most importantly,
having identified purity as a probability allows us to compute the \it
statistical \rm uncertainty associated with the purity of the sample.
Of course, in general there is also an additional associated
systematic uncertainty when one does not know the purity function with
infinite precision.

Turning our attention back to the expectation value for the number
density of clusters in a given bin, the relation $\avg{N|\Ns}=\Ns/p$
implies that the expected number density of clusters in a given
richness bin is given by
\begin{equation}
\avg{\na} = \int dm\ \nm \avg{c\psia/p|m}
\end{equation}
where 
\begin{equation}
\avg{c\psia/p|m} = \sum_{\Nobs,\Nt} P(\Nt|m)P_S(\Nobs|\Nt)\frac{c(\Nt)\psia(\Nobs)}{p(\Nobs)}
\end{equation}
and the sum extends over all $\Nt$ and $\Nobs$ values.

Before we end, we would like to reiterate our main point: the
importance of the above algebraic juggling is that, provided one is
willing to part with the information contained within the noise part
of the probability matrix, we can model observed cluster abundances
using only the signal matrix, the completeness function, and the
purity function.  Moreover, not only have we proved that we do not
need to know the details of the tails of the full probability matrix
$\pnn$, we have shown that not knowing these tails naturally gives
rise to the concepts of both completeness and purity.


\subsection{Photometric Redshift Uncertainties}
\label{sec:photozs}

An additional complication that we need to consider in our analysis is
the effect of photometric redshift estimation for the various
clusters.  Here, we make the simple assumption that photometric
redshift estimates can be characterized through a probability
distribution $\rho(\zc|\zh)d\zc$, where $\zh$ denotes the true halo
redshift and $\zc$ denotes the photometrically estimated cluster
redshift.  In general, we expect the probability $\rho(\zc|\zh)$ will
depend on the cluster richness $\Nobs$ since the number of galaxies
contributing to the photo-$z$ estimate for the cluster increases with
$\Nobs$.  On the other hand, systematic errors can mitigate the
sensitivity to cluster richness.  The bottom line is that when
applying our method to real data, it is important to check whether the
assumption that $\rho(\zc|\zh)$ is richness independent or not is a
valid one.  Generalizing to richness-dependent photometric redshift
errors is not particularly difficult.  We simply chose not to consider
this case in the interest of simplicity. Moreover, we will see later
that neglecting these dependences do not result in noticeable errors
in parameter estimation.

Consider then the expression for the total number of clusters in a
given richness bin and within some photometric redshift range
$[\zmin,\zmax]$.  We already know the comoving number density of halos
at redshift $\zh$ is given by equation \ref{eq:clden}.  To get the
number of clusters at an observed redshift $\zc$, we first multiply by
the comoving volume $(dV/d\zh)d\zh=A\chi^2(d\chi/d\zh)d\zh$ to get the
total number of clusters at redshift $\zh$, and then multiply by the
probability $\rho(\zc|\zh)d\zc$ that the clusters are observed within
some redshift range $d\zc$.  In the above expressions, $A$ is the area
of the survey and $\chi$ is the comoving distance to redshift $\zh$.
Summing over all halo redshifts, and over the photometric redshift
range considered $\zc\in[\zmin,\zmax]$, the total number of clusters
$\Na$ in a given bin is
\begin{equation}
\avg{\Na} = \int d\zh\ \int_{\zmin}^{\zmax} d\zc\avg{\na(\zh)} 
		\frac{dV}{d\zh} \rho(\zc|\zh).
\end{equation}

We will find convenient in the future  to rewrite the above expression in
terms of a redshift selection function $\varphi(\zc)$, defined to be unity if 
$\zc\in[\zmin,\zmax]$ and zero otherwise.  In terms of $\varphi$, the 
above expression becomes
\begin{equation}
\avg{\Na} = \int d\zh\ \avg{\na(\zh)} \frac{dV}{d\zh} \avg{\varphi|\zh}
\label{eq:clcts}
\end{equation}
where
\begin{equation}
\avg{\varphi|\zh} = \int d\zc\ \rho(\zc|\zh) \varphi(\zc).
\label{eq:zbin}
\end{equation}

The reason this recasting is useful is that in this language it
becomes obvious that the relation between $\zc$ and $\zh$ is the same
as that between $\Nobs$ and $\Nt$.  This then implies that when we set
out to compute the likelihood function for the observed number of
clusters $\Na$, the same algebra that describes uncertainties due to
richness estimation will describe uncertainties due to redshift
estimation, allowing us to quickly derive the relevant expressions for
one if we know the other.


\subsection{The Likelihood Function}
\label{sec:lkhd}

So far we have only concerned ourselves with developing a model for
the expectation value of the cluster number density.  In order to use
this model as a tool for extracting cosmological parameters from
observed cluster samples, we now attack the problem of modeling the
likelihood of observing a particular richness function.  In this work,
we chose to model the probability of observing a realization given a
set of cosmological and HOD parameters as a Gaussian.  While more
accurate likelihood functions can be found in the literature
\citep[][]{hucohn06,holder06}, these ignore correlations due to
scatter in the mass--observable relation, and thus we have opted for a
simple Gaussian model which is expected to hold if bins are
sufficiently wide (i.e. contain $\gtrsim 10$ clusters).  All we need
to do now is to compute the various elements of the correlation
matrix $\avg{\delta \na \delta\nap}$. Moreover, since our goal is to
perform a maximum-likelihood analysis, we calculate not the
correlation between cluster densities, but rather the correlation
matrix for the actual observed number of clusters in a given bin.  We
denote the number of clusters in richness bin $a$ as $\Na$, and the
correlation matrix as $C$, so that $\corr=\avg{\delta\Na\delta\Nap}$
where $\delta\Na=\Na-\avg{\Na}$.

The correlation matrix element $\corr$ has six distinct contributions:
\begin{enumerate}
\item A Poisson contribution due to the Poisson fluctuation in the
number of halos of mass $m$ within any given volume. 
\item A sample variance contribution reflecting
the fact that the survey volume may be slightly overdense or underdense with respect
to the universe at large.
\item A binning error arising from the stochasticity of $\Nobs$ as a function of $m$
and the probabilistic nature of the completeness function.
\item A contribution due to the statistical uncertainties associated with photometric 
redshift estimation.
\item A contribution due to the stochastic nature of the purity function.
\end{enumerate}

In principle, there is an additional pointing error for those cases in
which the central cluster galaxy is misidentified.  This error is
typically negligible except for small area surveys, or surveys with
highly irregular window functions, so we have opted to ignore this
effect.  We will see below that this did not affect our parameter
estimation in our mock catalogs at any noticeable level.  The
derivation of each of these contributions to the correlation matrix is
somewhat lengthy, so we shall simply refer the reader to
\citet[][]{hukravtsov03} and \citet[][]{rozoetal04}, which detail the
general procedure for deriving the relevant correlation matrices.  The
final expression for the various matrix elements can be found in
Appendix \ref{app:summary}.  Here, we simply take for granted that we
can compute correlation matrix $\corr$.  Given the correlation matrix
elements, and assuming flat priors for all of the relevant model
parameters, the likelihood function becomes
\begin{equation}
\lk(\bm{\Omega},\bm{p}|\bm{N}) = A\frac{\exp\left\{ - \frac{1}{2}(\bm{N} - \avg{\bm{N}}) 
	\cdot C^{-1}
	\cdot (\bm{N}-\avg{\bm{N}})\right\}}{\sqrt{(2\pi)^M \mbox{det}(C)}}	
\end{equation}
where $A$ is a normalization constant, 
$M$ is the number of bins, $\bm{N} = \{N_1,\ N_2,\ ...,\ N_M\}$
is the data vector, $\bm{\Omega}$ is the set of parameters describing cosmology and the 
HOD, and $\bm{p}$ is the set of nuisance parameters characterizing the purity
and completeness functions, and the parameters describing the signal matrix 
$\ps$. In general, calibration of the cluster-finding algorithm with simulations should
allow one to place strong priors on the distribution of the nuisance parameters,
in which case the above likelihood function is simply multiplied by the corresponding
\it a priori \rm (simulation-calibrated) probability distribution $\rho(\bm{p})$.


\section{Selection Function Calibration for the maxBCG Algorithm}
\label{sec:calibration}

In the previous section, we developed a general framework with which
one may quantitatively characterize the cluster selection function of
any cluster finding algorithm.  We now proceed to calibrate this
selection function for the maxBCG cluster finding algorithm from
\citet[][]{koesteretal06a}.  The idea behind the calibration is
simple: we use an empirically driven algorithm to populate dark matter
simulations with galaxies, resulting in realistic galaxy catalogs
comparable to the SDSS data.  We then run the maxBCG cluster-finding
algorithm on each of our mock galaxy catalogs, and compare the
resulting mock cluster catalogs to the original input halo catalog of
the simulation to directly measure the matrix elements $\pnn$.  In
what follows, we only briefly describe both the mock catalogs and the
cluster finding algorithm, as the relevant details can be found
elsewhere.  Rather, we focus on the key aspects of the analysis that
are particular to the general framework developed earlier in section
\ref{sec:model}.


\subsection{The Simulations}

The N-body simulation based mock catalogs we use in our calibration
are described in detail in \citet[][]{wechsleretal07}.  Briefly,
galaxies are attached to dark matter particles in the Hubble Volume
light-cone simulation described in \citet[][]{evrardetal02} using an
observationally-motivated biasing scheme.  The relation between dark
matter particles of a given overdensity (on a mass scale of $\sim 1e13
M_{\odot}$) is related to the correlation function of the particles;
these particles are then chosen to reproduce the luminosity-dependent
correlation function as measured in the SDSS by
\citet[][]{zehavietal05}.  The number of galaxies of a given
brightness placed within the simulations is determined by drawing
galaxies from the SDSS galaxy luminosity function
\citep[][]{blantonetal03}.  Finally, colors are assigned to each
galaxy by measuring their local galaxy density, and assigning to them
the colors of a real SDSS galaxy with similar luminosity and local
density (see also \citealt{tasitsiomietal04}).  This method produces
mock galaxy catalogs that reproduce several properties
of the observed SDSS galaxies, and in particular follow the empirical
color-galaxy density relation and its evolution, which is
is particularly important for ridgeline based cluster detection methods.  In this
work, we use three different realizations of this general populations
scheme.  The resulting catalogs are labeled Mocks A, B, and C.  Each of
these catalogs has different HOD, which allows us test the robustness
of the selection functions to varying cosmologies.


\subsection{The maxBCG Cluster-Finding Algorithm}

Details of how the maxBCG cluster-finding algorithm works can be found
in \citet[][]{koesteretal06a}.  Briefly, maxBCG assumes that the
Brightest Cluster Galaxy (BCG) in every cluster resides at the center
of the cluster.  These BCG galaxies are found to have a very tight
color-magnitude relation, which is used to select candidate BCGs, and
to evaluate the likelihood $\lk_{BCG}$ that these candidates are
indeed true BCGs.  In addition, maxBCG uses the fact that all known
clusters have a so-called ridgeline population of
galaxies, bright early-type galaxies that populate a narrow ridgeline
in color--magnitude space.  Using a model for the radial and color
distribution of ridgeline galaxies in clusters, the likelihood $\lk_R$
that the galaxy population around the candidate BCGs is due to a
cluster being present is computed.  These likelihoods are maximized as
a function of redshift, which provides a photometric redshift estimate
for the cluster.  The candidate BCGs are then rank ordered according
to the total likelihood $\lk=\lk_R\lk_{BCG}$.  The top candidate BCG
is selected as a cluster BCG, and its satellite galaxies are removed
from the candidate BCGs.  In the algorithm, all ridgeline galaxies
within a specified scaled radius $R_{200}$ of the cluster are
considered satellite galaxies (details are given in
\citet[][]{koesteretal06a}).  The process is then iterated until a
final cluster catalog is obtained.


\subsection{Matching Halos to Clusters}
\label{sec:matching}

Given a halo catalog and the corresponding mock cluster catalog,
estimating the matrix element $P(\Nobs|\Nt)$ becomes a simple matter
of measuring the fraction of halos with $\Nt$ galaxies detected as
clusters with $\Nobs$ galaxies.  Of course, in order to compute this
fraction, one needs first to define $\Nt$ and $\Nobs$, and then one
needs to know how to find the correct cluster match for individual
halos in the halo catalog.  Concerning the first point, and in the
interest of having a volume limited catalog to $z=0.3$, we define
$\Nt$ as the number of galaxies in a halo (i.e. within $r_{200}$,
where $r_{200}$ is the radius at which a the mean density of the
cluster is $200$ times the critical density of the universe) above an
$i$-band luminosity of $0.4L_*$.  Note that no color cut is applied in
the definition of $\Nt$.  As mentioned earlier, we take
$\Nobs$ to be simply the $N_{gals}^{r200}$ richness estimate from
\citet[][]{koesteretal06a}.  Note that in general, the probability
matrix $P(\Nobs|N_{\rm T})$ will depend on precisely how one defines
galaxy membership for both halos and clusters.  In this work, we focus
exclusively on the above definitions, and leave the problem of whether
our result can be improved upon by a redefinition of halo and cluster
richness for future work.  The above definitions are intuitively
reasonable ones, and thus provide a good starting point for our
analysis.


\begin{figure}[t]
\epsscale{1.2}
\plotone{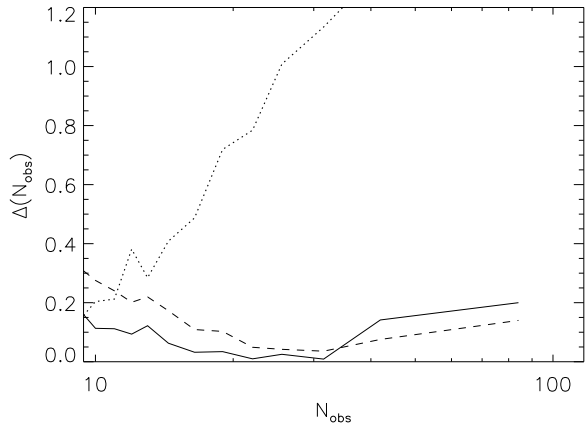}
\caption[Cost Function for Halo-Cluster Matchings]{The cost function
  $\Delta(\Nobs)$ defined in equation \ref{eq:cost}. The function
  compares the observed abundance to that predicted using the matching
  between observed and halo-based richness, ignoring purity and
  completeness.  The solid line is obtained using exclusive maximum
  shared membership matching to compute the probability matrix
  estimate $\hat \pnn$. The dashed line is obtained with exclusive BCG
  matching, while the thin dotted line is obtained using non-exclusive
  maximum membership matching.  Other non-exclusive matchings,
  including the probabilistic algorithms discussed in the text, look
  quite similar to the dotted line.  The upturn at high richness for
  the one to one matchings are due to catastrophic errors in the
  cluster-finding algorithm (i.e. noise term in the $\hat \pnn$
  matrix) and are unphysical. We chose exclusive maximum shared
  membership matching as our fiducial halo-cluster matching
  algorithm.}
\label{fig:matching}
\end{figure} 


We now turn to the problem of matching halos to clusters.  In general,
there is no unique way of matching halos to clusters and vice-versa.
For instance, a halo could be matched to the cluster that is found
nearest to it, to the halo that contains the clusters central galaxy,
or to the cluster that contains the largest fraction of that halo's
galaxy members.  Note that since different matching schemes will
result in different probability matrices, one needs to consider
multiple schemes and determine which is most correct.  We define a
matching algorithm to be optimal if it minimizes the cost function
\begin{equation}
\Delta(\Nobs) = \frac{|n(\Nobs)-\sum_{\Nt} n(\Nt) \pnn|}{n(\Nobs)}.
\label{eq:cost}
\end{equation}
Note that if matching was perfect, we would expect $\Delta=0$.  The
cost function is closely related to the purity and completeness
function.  In particular, the cost function is a measure of how
strongly the observed cluster sample deviates from having completeness
and purity exactly equal to unity in the absence of pointing and
photometric redshift errors.  As detailed in Appendix
\ref{app:matchmodels}, we find that the matching algorithm that
worked best was one in which the richest halo is matched to the
cluster with which it shares the largest number of galaxies.  The halo
and cluster are then removed from the halo and cluster catalogs
respectively, and the procedure is iterated.  Figure
\ref{fig:matching} demonstrates that the resulting cost function is
below $20\%$ at all richnesses when using this particular matching
scheme, which immediately tells us that the purity and completeness of
the sample are better than $80\%$.


\subsection{Signal and Noise}
\label{sec:signal_and_noise}


\begin{figure}[t]
\epsscale{1.2}
\plotone{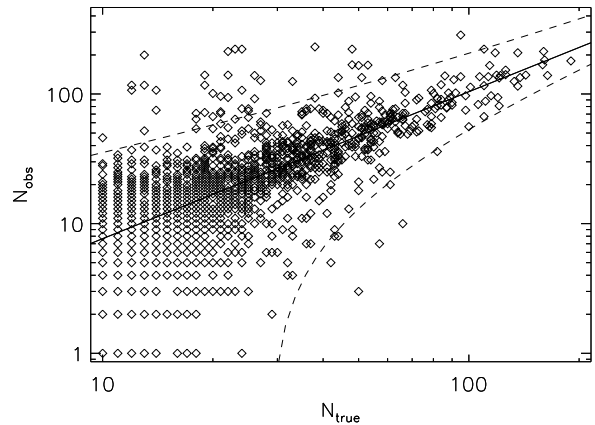}
\caption[The Probability Matrix and the Signal/Noise
Decomposition]{The estimated probability matrix $\hat \pnn$ in Mock A.
  Non-zero matrix elements are marked with diamonds.  The best-fit
  power law to the maximum-likelihood relation between $\Nobs$ and
  $\Nt$ is shown above as the thick solid line.  The dashed curves
  define the signal band: everything within these lines is considered
  signal in the sense that it corresponds to proper halo-cluster
  matches.  Points outside this band, including the points on each
  axis, are considered noise in the sense that they represent
  catastrophic errors of the cluster-finding algorithm (see text for
  how the dashed lines are defined); these points will contribute to
  the incompleteness and the impurity of the sample.  Note that
  blending, that is, matching of low richness halos to high richness
  clusters, is clearly more of a problem than halo splitting (matching
  of high richness halos to low richness cluster), as argued in
  \citet[][]{koesteretal06a}.  We only show $\Nt\geq 10$ as this
  corresponds to the resolution limit of the simulation.}
\label{fig:pnn}
\end{figure} 


The observed fraction $\hat P(\Nobs|\Nt)$ of halos with $\Nt$ galaxies
that are matched to clusters with $\Nobs$ galaxies is our estimator
for the matrix element $P(\Nobs|\Nt)$.\footnote{When performing the
  matching, it is important to keep in mind that the halo catalog
  should have a slightly smaller area and redshift range than the
  corresponding cluster catalog.  This is because due to pointing and
  photometric redshift errors a halo located near the boundary of the
  survey could be well matched by a cluster just outside said
  boundary.}  Figure \ref{fig:pnn} shows the non-zero matrix elements
of the estimated probability matrix.  We can see that this plot has
the generic behavior we expected from \S \ref{sec:model}: the
majority of the non-zero matrix elements populate a diagonal band,
with a few outliers which arise due to catastrophic errors in the
cluster finding algorithm.  We split the matrix $\pnn$ into a signal
and a noise component as follows: first, we find the
maximum-likelihood relation between $\Nobs$ and $\Nt$, that is, we
select the matrix elements that maximize $\pnn$ at fixed $\Nt$,
restricting ourselves to the region $\Nt\geq 10$ as this corresponds
to the resolution limit of the simulations (ie, some halos which would
contribute to lower $\Nt$ could be missed).  The maximum-likelihood
matrix elements are then fit with a line $\mu(\Nt)$ using robust
linear fitting \citep[][]{numrecipes}.  The best-fit line is
characterized by the two parameters $B_0$ and $\beta$ defined via
\begin{equation}
\mu_{ML}(\Nt) = 20\exp(B_0)(\Nt/20)^{\beta}.
\label{eq:muML}
\end{equation}

We now make
the ansatz that the variance will, at least roughly, scale with this maximum
likelihood relation, which would be the case for a Poisson-like process.  The
signal band is then defined as all matrix elements $(\Nt,\Nobs)$ such that
\begin{eqnarray}
\Nobs & \geq & \mu_{ML}(\Nt)-5\sqrt{\mu_{ML}(\Nt)},\ \mbox{and} \\
\Nobs & \leq & \mu_{ML}(\Nt)+10\sqrt{\mu_{ML}(\Nt)}
\end{eqnarray}
Non-signal halo-cluster pairs are defined to be noise.  The
signal/noise decomposition for Mock A can be seen in Figure
\ref{fig:pnn}: any matrix elements contained within the dashed lines
are signal, and everything outside the dashed lines constitutes noise.
The solid line going through the signal band is our best fit of the
maximum-likelihood relation between $\Nobs$ and $\Nt$.  Plots for the
probability matrix for the other two mocks are qualitatively very
similar.

While the above procedure is ad-hoc, we emphasize that we are \it
defining \rm the signal band.  At the end of the day, it does not
matter how we came up with the above definition, what matters is
whether the definition is a useful one or not.  For our purposes, we
have simply selected a straightforward algorithm that qualitatively
does what we need it to do, that is, cleanly separate signal points
from catastrophic errors.  Other algorithms would certainly be
possible and equally valid, and there will undoubtedly be a definition
that works best in the sense that that the statistical errors from a
maximum likelihood analysis using the likelihood from \S
\ref{sec:model} would be minimized.  In this work, we simply wish to
adopt a working definition, and we demonstrate below that even with
this simplest definition where signal and noise are separated by eye,
our model likelihood correctly describes the data and we can
successfully recover the cosmological and bias parameters with exquisite
precision.


\subsection{Completeness}

\begin{figure}[t]
\epsscale{1.2}
\plotone{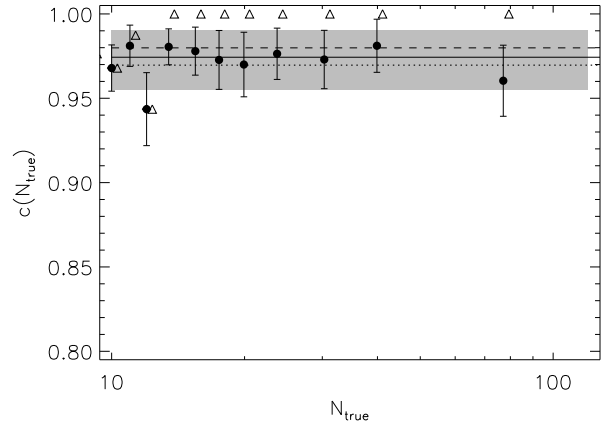}
\caption[The maxBCG Completeness Function]{The completeness function
  $c(\Nt)$ as measured in Mock A.  Filled solid circles with error
  bars are the observed fraction of halos of richness $\Nt$ matched
  within the signal band in Figure \ref{fig:pnn}. For comparison, we
  also show the fraction of halos matched to clusters of any richness
  as triangles.  The best fit completeness, modeled as constant, is
  shown as a solid line, and the grey regions represents the $95\%$
  confidence interval in our completeness determination.  Dashed and
  solid lines are obtained for Mocks B and C
  respectively, which are fully consistent with each other.}
\label{fig:completeness}
\end{figure}

Having separated our halo-cluster pairs into a signal and a noise
component, our estimator $\hat c(\Nt)$ for the completeness function
is simply the fraction of halos with $\Nt$ that are considered signal.
In order to reduce the noise in these estimators, we also bin our
halos into richness bins by demanding that each richness bin contain
at least 50 halos.  The resulting estimated completeness function in
Mock A is shown in Figure \ref{fig:completeness} as the solid circles
with error bars. Also shown in Figure \ref{fig:completeness} as
triangles is the fraction of halos matched to cluster of any richness,
regardless of whether the match constitutes signal or noise.  We can
see that for relatively rich systems with $\Nt\gtrsim 25$, essentially
all halos are detected, but completeness differs from unity due to
some of these halos being blended.  Conversely, at low richness,
the vast majority of detected halos constitute signal, but the
detected fraction decreases with decreasing richness, and as a result,
the completeness function is essentially flat.  We found this to be
the case in each of the mocks.

We model the completeness function as a constant independent of
richness.  Our best-fit model for the completeness function is defined
via $\chi^2$ minimization, and is seen in Figure
\ref{fig:completeness} as a thick, solid line.  Best fits for Mocks B
and C are also shown as dashed and dotted lines respectively.  Error
bars for the $\chi^2$ minimization are assigned using the fact that
the number of signal halos follows a binomial distribution with a
detection probability $c(\Nt)$, from which we can compute the expected
standard deviation of the ratio of signal halos to all halos in a
given richness bin.


In order to determine whether our $\chi^2$ fit is a good fit, and to
estimate the uncertainty in the best-fit completeness, we performed
$10^4$ Monte Carlo realizations of our best-fit completeness model,
and then treated these realizations in the same way we treated our
data. We found the $\chi^2$ values measured in the mock to be
consistent with our Monte Carlo $\chi^2$ distribution. The $95\%$
confidence interval for the completeness function in Mock A is shown
in Figure \ref{fig:completeness} as a grey band.  Each of the mocks
have completeness measures that are consistent with each other.


\subsection{Calibrating the Signal Matrix}
\label{sec:signalmatrix}

Calibration of the signal matrix $\ps$ is essentially a trial and error game
since we have no a priori expectation for the form of $\ps$.  Note, however,
that so long as we parameterize $\ps$ in a way that is statistically consistent
with the simulation constraints, it does not matter how we came up with
our particular parameterization.  Rather than discussing the various iterations we
went through to find a successful model for $\ps$, we have chosen to 
simply state our model, and then demonstrate that our parameterization 
is flexible enough to fully accommodate the simulation data.
Our model for the probability matrix is 
\begin{equation}
\ps \propto \mbox{\rm erf}(x_{max})-\mbox{\rm erf}(x_{min})
\label{eq:prob1}
\end{equation}
where $x_{min}=(\Nobs-\mu(\Nt)+1)/\sqrt{2\mbox{Var}(\Nt)}$, 
$x_{max}=(\Nobs-\mu(\Nt)+2)/\sqrt{2\mbox{Var}(\Nt)}$, and
\begin{eqnarray}
\mu(\Nt) & = & 20\exp(B_0+0.14)(\Nt/20)^{(\beta-0.12)} \\
\mbox{Var}(\Nobs|\Nt) & = & \exp(-3B_0+B_1)\mu(\Nt).
\end{eqnarray}
The factor of $20$ is simply our chosen pivot point.  Note that $B_0$
and $\beta$ were defined earlier in equation \ref{eq:muML}, so that
the only new parameter being introduced is $B_1$, which characterizes the
variance of $\ps$.  The appearance of $-3B_0$ in the expression for
Var$(\Nt)$ de-correlates $B_0$ and $B_1$, whereas the additive
constants $0.14$ and $-0.12$ in the expressions for $\mu(\Nt)$ were
empirically determined and characterize the difference between the
mean value $\avg{\Nobs|\Nt}=\mu(\Nt)$ and the maximum likelihood value
$\mu_{ML}(\Nt)$ from equation \ref{eq:muML}.  Finally, the
proportionality constant in equation \ref{eq:prob1} is set by
demanding that the sum of all matrix elements over the signal band be
equal to unity.


\begin{figure}[t]
\epsscale{1.2}
\plotone{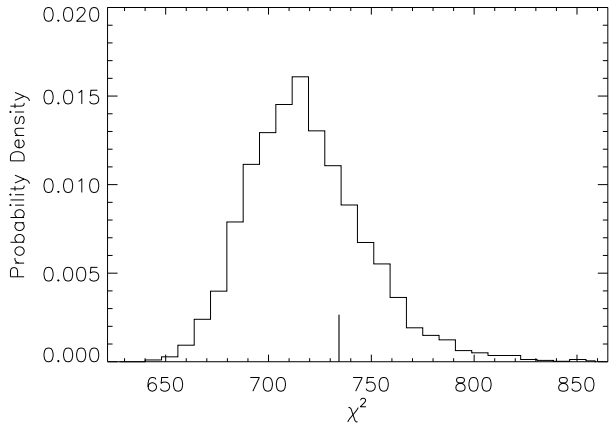}
\caption[$\chi^2$ Distribution for the Model Signal Matrix and the Simulation Value]{$\chi^2$ 
distribution of $10^4$ Monte Carlo realizations of our best-fit
model of the signal matrix $\ps$ for Mock A.  The $\chi^2$ value 
observed in the mock is marked by a thick, solid line at the bottom of
the plot.  We find that our model for $\ps$ is indeed a good fit to the mock catalog
data.  This was the case for Mocks B and C as well.}
\label{fig:spschi}
\end{figure} 


We now demonstrate that this parameterization does indeed provide a
good fit to the mock catalogs and estimate the uncertainties in our best
fit parameters.  To do so, we first find the best-fit value for $B_1$
by minimizing $\chi^2$ and assuming a binomial distribution for
computing the error bars for each matrix element.  We then compare the
$\chi^2$ distributions obtained from $10^4$ Monte Carlo realizations
of our best-fit model for each of our mocks to the $\chi^2$ value
observed in the mocks directly. Figure \ref{fig:spschi} illustrates
our result in the case of Mock A. It is clear from the
figure that the model is indeed a good fit to the simulation data.
This is true of Mocks B and C as well.\footnote{An exact
  comparison between our Monte Carlo realizations and the mocks
  suffers from the fact that while the mocks suffer from
  completeness being different from unity, whereas our Monte Carlo
  models of the probability matrix do not.  Since $\ps=\pnn/c(\Nt)$,
  there is somewhat of an ambiguity as to whether in comparing the two
  concerning whether we should inflate the error estimates of the
  Monte Carlo realizations by a factor of $1/c$ as we do for the
  simulation data.  Fortunately, since the completeness is close to
  unity, this ambiguity does not alter the $\chi^2$ distributions
  much (we checked this explicitly).}

\begin{figure}[t]
\begin{center}
\epsscale{2.4}
\plottwo{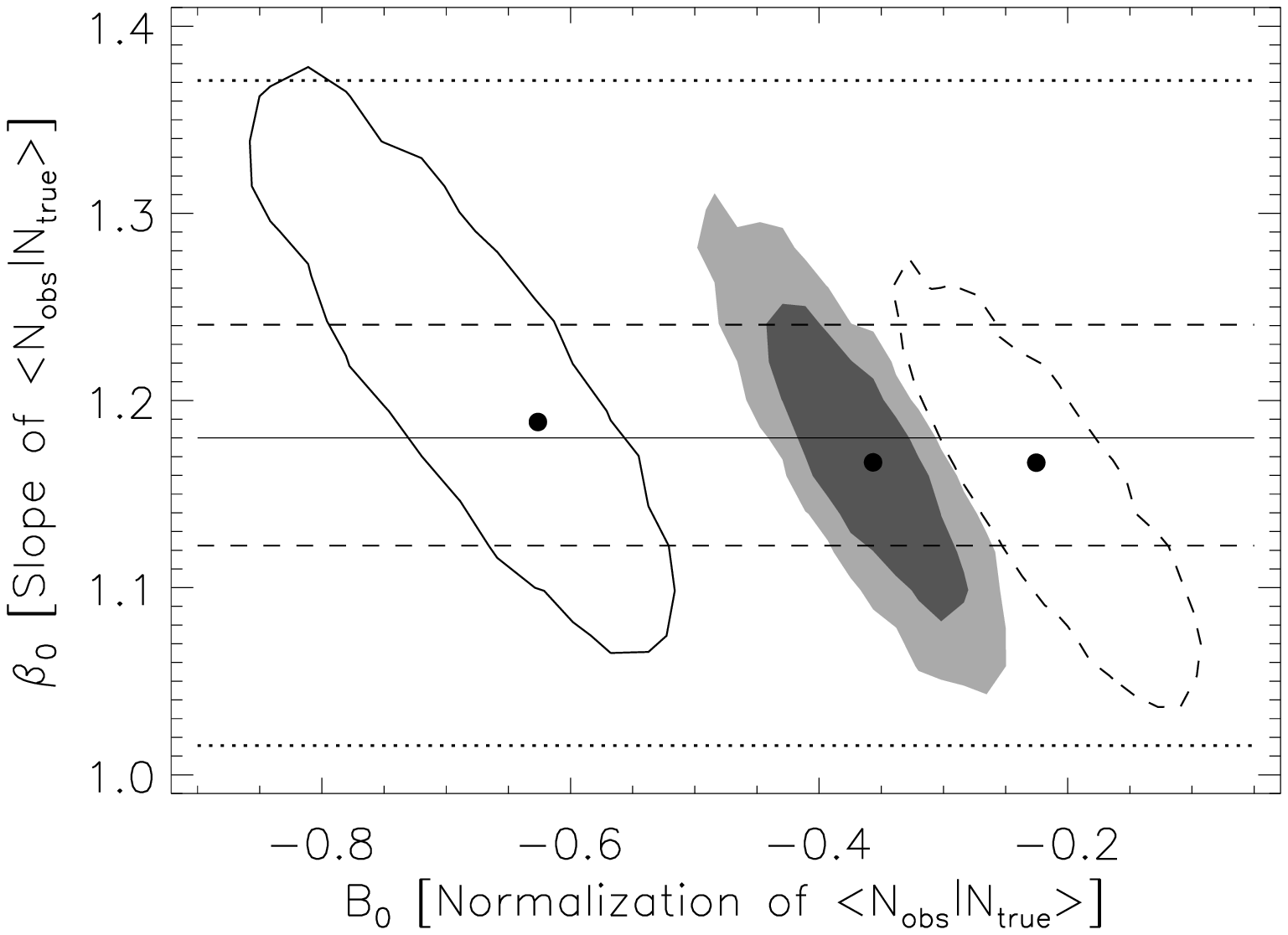}{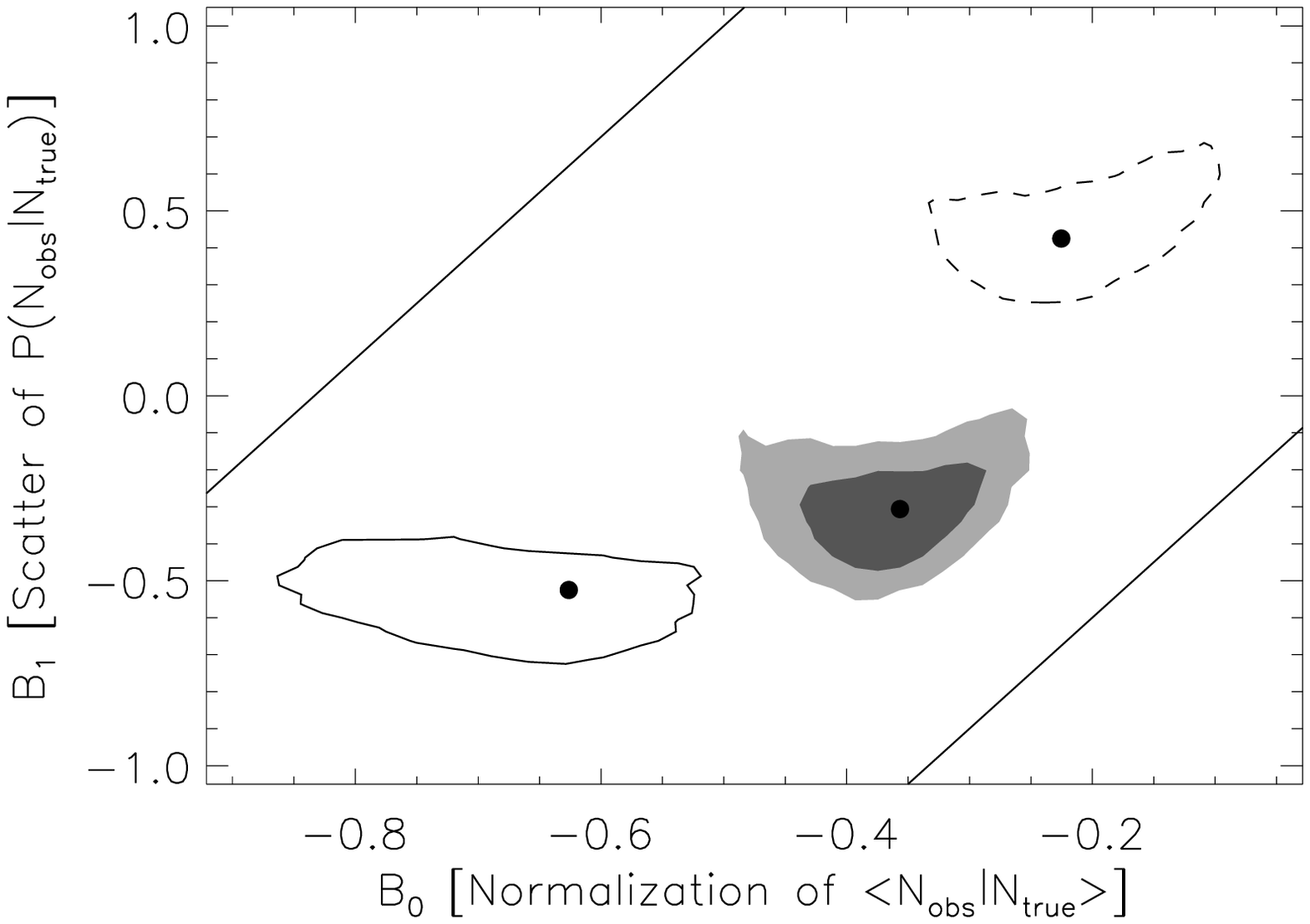}
\end{center}
\caption[Confidence Regions of the Signal Matrix Parameters]{$95\%$
  confidence regions for the $B_0$ and $\beta$ parameters (top) and
  $B_0$ and $B_1$ (bottom) in each of our three mock catalogs.  The
  dashed and solid contours are for Mocks A and B respectively.  The
  shaded contours are $68\%$ and $95\%$ confidence regions in Mock
  A. The small filled circles mark the best-fit parameters from the
  mock catalogs, and were used to generate the Monte-Carlo realizations
  from which the confidence regions are derived.  With the exception
  of $\beta$, the best-fit parameters in each of the mocks are
  clearly not consistent with each other, and represent a large
  systematic error.  The solid line in the top panel corresponds to
  the value $\beta=1.18$, while the dashed lines mark the assumed
  1$\sigma$ error $\Delta\beta/\beta=5\%$ in this work.  The dotted lines 
  in the top panel and the solid lines in the bottom panel mark the 
  assumed $1\sigma$  region used in \citet[][]{rozoetal07b}. }
\label{fig:compareps}
\end{figure}

The top plot in Figure \ref{fig:compareps} shows the $95\%$ confidence
regions of the parameters $B_0$ and $\beta$ in Mocks A, B, and C.  The
corresponding regions for the parameters $B_0$ and $B_1$ are shown in
the bottom panel.  It is evident that the best-fit parameters $B_0$
and $B_1$ in each mock are not fully consistent with each other, and
that this variation represents a large systematic uncertainty in
the probability matrix $\ps$.  Nevertheless, the slope $\beta$ appears
to be robustly constrained, with roughly $\beta=1.18\pm 5\%\ (1\sigma)$.  
Early analysis of some recent simulations suggests that the scatter in $\beta$
is in fact larger than this, and the mean is somewhat lower, around $\beta\approx 1.1$,
though a complete study of these simulations has not yet been completed.
In what follows, we simply assume $\beta=1.18\pm 5\%$ unless noted otherwise,
though we note we use a much more conservative prior $\beta=1.18\pm 15\%$ 
when analyzing the maxBCG cluster catalog constructed from SDSS data.


\subsection{Purity}

\begin{figure}[t]
\epsscale{1.2}
\plotone{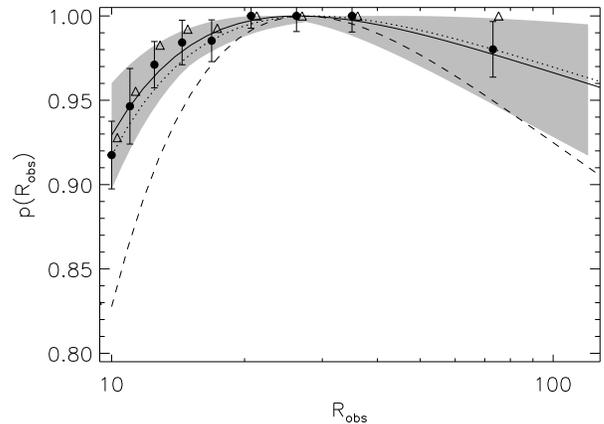}
\caption[The maxBCG Purity Function]{Fraction of clusters matched to a halo within the signal
band.  Filled circles are the fraction measured in Mock A.  The thick
solid line is the best fit to the data, and the grey band represents the
$95\%$ confidence band of the model.  Dotted and dashed lines are the best fits
for Mocks B and C respectively.  For reference, we also
show with triangles 
the fraction of clusters matched to halos of \it any \rm richness.}
\label{fig:purity}
\end{figure}

The purity function represents the fraction of clusters that are not
well matched to a halo (i.e. that fall outside the signal band).
Calibration of the purity function is thus completely analogous to the
completeness function provided we switch the role of halos and
clusters.  That is, thinking of clusters as input and halos as output,
we follow the exact same procedure we used to define the completeness
function in order to define the purity function.  Our estimate for of
the purity function in Mock A is shown in Figure \ref{fig:purity}.
Also shown as a solid line is our best-fit model, which we have chosen
to parameterize as
\begin{equation}
p(\Nobs) = \exp(-x(\Nobs)^2)
\end{equation}
where
\begin{equation}
x(\Nobs)=p_0+p_1\left(\frac{\ln(15)}{\ln(\Nobs)}-1\right).
\end{equation}
The factor of $\ln(15)$ is there simply to de-correlate $p_0$ and
$p_1$.  The best-fit models for Mocks B and C are also shown as dashed
and dotted lines respectively.  Finally, as with completeness, we
generate $10^4$ Monte Carlo realizations to estimate the $95\%$
confidence regions of our best-fit parameters, and to test whether the
model is a statistically acceptable fit to the data.  We find that
this is indeed the case, and the corresponding $95\%$ confidence band
for Mock A is shown in Figure \ref{fig:purity} as a grey band.  The
three mock catalogs are only marginally consistent with each other,
but the purity is quite high in each of them.


\subsection{Photometric Errors Calibration} 

We now calibrate the probability distribution $\rho(\zc|\zh)$,
that is, the distribution of photometric cluster redshift estimates in terms
of the true halo cluster redshift.  We characterize the photometric
redshift distribution in terms of the redshift bias
parameter $b=\zc/\zh$.  The probability distribution $\rho_b(b|\zh)$ is related
to the probability distribution $\rho(\zc|\zh)$ via 
\begin{equation}
\rho(\zc|\zh) = \frac{1}{\zh}\rho_b(b|\zh).
\label{eq:probrel}
\end{equation}

The advantage of working with $\rho_b(b)$ is that \it b correlates
only very weakly with halo redshift $\zh$. \rm Indeed, we found that
the cross correlation coefficient between $b$ and $\zh$ in our mock
catalogs was $\lesssim 0.1$.  Moreover, we found the cross correlation
between $b$ and $\Nobs$ to be equally weak, so taking $\rho_b(b)$ to
be richness independent is a good approximation for the maxBCG cluster
catalog at richness $\Nobs\geq 10$ (the richness range that will be
used for cosmological constraints).

\begin{figure}[t]
\epsscale{1.2}
\plotone{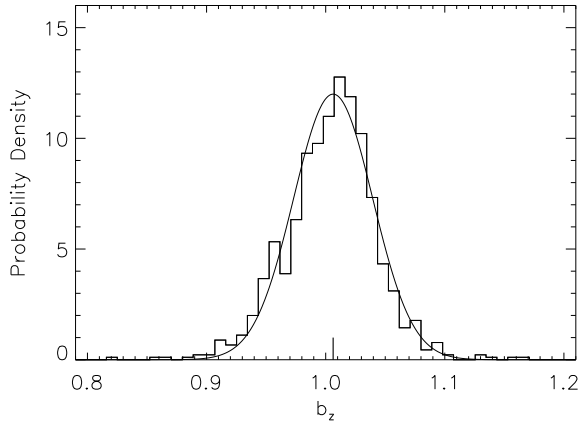}
\caption[The maxBCG Photometric Redshift Bias Distribution]{
  Distribution of the photometric redshift bias parameter $b=\zc/\zh$
  for Mock A.  The distribution $\rho_b(b)$ is richness and redshift
  independent to a good approximation, and is well fit by a Gaussian,
  as shown above.  The thick solid line at the bottom represent the
  best-fit value for the average bias.}
\label{fig:photozs}
\end{figure}

Figure \ref{fig:photozs} shows the distribution of bias parameters for
each halo--cluster pair in Mock A.  The distribution $\rho(b)$ is seen
to be well fit by a Gaussian, and is thus completely characterized by
the average bias parameter $\avg{b}$ and its standard deviation
$\sigma_b$.  The best-fit parameters for each of our mocks are
$\avg{b}=1.00,1.02,$ and $1.03$ and $\sigma_b=0.04,0.03,$ and $0.05$
for Mocks A, B, and C respectively.  These determinations have
effectively zero statistical error; here again
systematic variations from realization to realization represent
our main source of uncertainty.

%
%


\section{Testing the Model}
\label{sec:model_testing}

We now test whether our model can successfully reproduce the observed
number counts in the mock catalogs.  Even more importantly, we test
whether we can successfully recover the cosmological and HOD
parameters of the simulations with use of the likelihood function from
\S \ref{sec:model}.  Throughout this section, we use the
\citet[][]{jenkinsetal01} parameterization of the halo mass function.
The linear power spectrum is computed using the low baryon transfer
functions from \citet[][]{eisensteinhu99} with zero neutrino masses,
and the initial power spectrum is assumed to be a Harrison-Zeldovich
spectrum.  Flatness is also assumed, and all cosmological
parameters are held fixed except for $\sigma_8$, $\Omega_m$, and $h$.
Allowing other parameters to vary should have only a minor impact on our
results as it has been shown \citep[][]{whiteetal93,rozoetal04} that
local halo abundances are most sensitive to these three parameters.

\subsection{Cluster Counts Comparison}
\label{sec:counts_comparison}

We begin by comparing the cluster number counts in each of our three
mocks to our model predictions using the simulation-calibrated values
for all eight nuisance parameters (one completeness, two purity, two
photo-$z$, and three signal matrix parameters). The input cosmology
and HOD are taken directly from the mock catalog.  There are, however,
two important points concerning the mocks which we would like to
highlight.  First, the variance in the number of galaxies in halos of
a given mass is somewhat larger than Poisson.  Consequently, in this
section we take $P(\Nt|m)$ to be Gaussian, and calibrate the relation
between $\mbox{Var}(\Nt|m)$ and $\avg{\Nt|m}$ directly from the mocks.
Secondly, halo masses in the simulation were defined at an overdensity
$\Delta=200$ with respect to critical, whereas our model requires masses to be
measured at an overdensity of $200$ relative to the mean matter density.
We transform the masses accordingly using the fitting functions from
\citet[][]{hukravtsov03}.  Our final uncertainty in the fitted values
for the HOD is $\approx 3\%$ as estimated by examining the sensitivity
of our best-fit parameters to the number of bins used for the
calibration and the minimum mass cut considered when fitting the HOD.
We will see below that this accuracy is comparable to the statistical
uncertainty with which we can recover the best-constrained modes in
parameter space.


\begin{figure}[t]
\epsscale{1.2}
\plotone{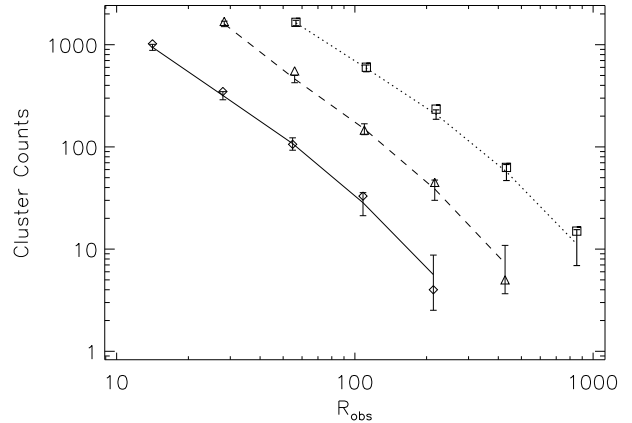}
\caption[Comparison of Simulated Cluster Counts With Predicted
Counts]{Comparison between the cluster counts measured in the mocks
  and our model predictions.  Solid, dotted, and dashed curves
  represent Mocks A, B, and C.  For clarity, we have also displaced
  mocks C and B to the right by a factor of $2$ and $4$ respectively.
  Error bars on the model values are obtained from the diagonal terms
  of the correlation matrix, and are roughly uncorrelated.}
\label{fig:simcounts}
\end{figure} 


Figure \ref{fig:simcounts} shows the cluster number counts in each of
our three mock catalogs as well as our model predictions.  The error
bars associated with the model are simply the square root of the
diagonal terms in the correlation matrix, and we have selected bins
wide enough for the error bars to be roughly de-correlated.  The
agreement between the model predictions and the observed number counts
in the mocks is excellent.  Note that this agreement is \it not \rm
trivial.  While it is true that our mass function is calibrated to the
simulations, agreement between our prediction and the direct
measurement in the mocks is only assured if our model successfully
parameterizes the cluster-finding algorithm selection function.
Figure \ref{fig:simcounts} demonstrates that this is indeed the case,
and that any systematics in the data have been properly taken into
account.


\subsection{Parameter Constraints for a Known Selection Function}
\label{sec:fixed_nuisance}

We wish to test now whether we can successfully recover the
cosmological and HOD parameters of the simulation with the use of the
likelihood function constructed in \S \ref{sec:model}.  To do so, we
use a Monte Carlo Markov Chain (MCMC) method to evaluate the
likelihood function in parameter space and estimate the corresponding
$68\%$ and $95\%$ confidence confidence contours of the likelihood
function in parameter space.  Details of our MCMC implementation,
which draws heavily on the work by \citet[][]{dunkleyetal05}, can be
found in Appendix \ref{sec:implementation}.  Throughout this paper, we
consider cluster number counts binned in nine logarithmic bins between
richness $\Nobs=10$ and $\Nobs=100$ within a single redshift slice
($[z_{min},z_{max}]=[0.12,0.25]$.  We chose this
binning as a compromise between having enough bins to accurately
resolve the shape of the cluster richness function, while at the same
time ensuring that every bin contained $\gtrsim 10$ clusters.  This
last property is desirable since the Gaussianity assumption of our
likelihood function breaks down if the number of cluster within a
given bin becomes too low.  

We begin our analyzes by estimating the likelihood function while
holding all of our nuisance parameters fixed to the simulation-
calibrated values.  That is, we assume we have perfect knowledge of
the cluster selection function.  This is useful for two reasons:
first, it allows us to test whether our model likelihood successfully
recovers the input cosmology and HOD parameters when the cluster
selection function, that is, the probability matrix $\pnn$, is fully
calibrated.  In addition, investigating this case gives gives us a
baseline for evaluating how well the signal matrix parameters must be
calibrated before the quality of our parameter constraints decreases
significantly.  We should also note that, by and large, holding the
cluster selection function fixed is a standard assumption in most
analyses of cluster abundances.  One of the most powerful features of
our method is that, as we shall see in \S \ref{sec:float_nuisance}, it
allows us to marginalize over uncertainties in the selection function.


\begin{figure}[t]
\epsscale{1.2}
\plotone{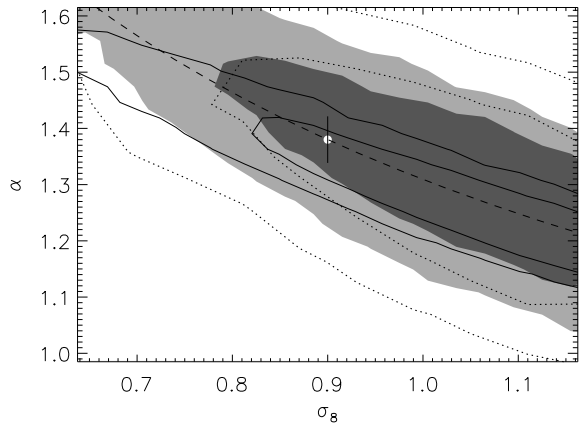}
\caption[Confidence Regions in the $\sigma_8-\alpha$ Plane Observed in Simulations]{Filled 
contours are the $68\%$ and $95\%$ confidence regions in the 
$\alpha-\sigma_8$ plane
recovered for Mock A when holding all nuisance parameters fixed.
The true parameters are marked as a small circle with error bars.  We find that
our likelihood model successfully recovers the simulation parameters to within
the degeneracies intrinsic to the data.   
Also shown with solid curves are the $68\%$ and $96\%$ confidence 
regions obtained when using CMB and supernova like Gaussian priors 
$\Delta\Omega_mh^2=0.01$ and $\Delta h=0.05$.  The dotted contours
are obtained by marginalizing over all completeness, purity, and photo-$z$
parameters, and assuming $10\%$ priors on the signal matrix parameters (see
\S \ref{sec:float_nuisance} for discussion).  The dashed line marks the expected
degeneracy direction $\alpha^2\sigma_8=constant$, while the error bar
centered on the true values of the simulation represents the $3\%$ error on
$\alpha$ from the direct measurement of the HOD.
Results for the three mocks considered were all very similar.}
\label{fig:ssim_s8_aN}
\end{figure} 


As expected, we find that there are strong degeneracies between
between cosmology and HOD parameters.  This is illustrated in Figure
\ref{fig:ssim_s8_aN}, where we plot the $68\%$ and $95\%$ confidence
regions in the $\alpha-\sigma_8$ plane for Mock A.  All three mocks
give similar results.  The degenerate parameter combination is roughly
$\alpha^2\sigma_8 = constant$, in agreement with the Fisher matrix
estimate from \citet[][]{rozoetal04}.  Also shown in this figure with
a small circle with error bars are the known value of the parameters
in the simulation; the error bars represent the $\approx 3\%$
uncertainty in our direct measurement of the HOD parameters.  Clearly,
to within the degeneracies intrinsic to the method, we successfully
recover the input cosmology and HOD.

In light of the strong degeneracies inherent to the data, we focus now
on the directions in parameter space that are best constrained by the
data.  These are defined by diagonalizing the parameter correlation
matrix as estimated from the MCMC output.  The best-constrained modes
are those for which the eigenvalues are smallest.  In the case of the
Mock A, the top two normal modes are
\begin{eqnarray}
x_1 & = & \alpha^{0.97}\sigma_8^{0.92}(\Omega_m/M_1)^{0.35}(\Omega_mM_1)^{-0.06}h^{-0.45} \\
x_2 & = & \alpha^{1.60}\sigma_8^{-0.26}(\Omega_m/M_1)^{0.54}(\Omega_mM_1)^{0.25}h^{0.12}.
\end{eqnarray}
Note the first parameter is essentially a cluster normalization
condition, but with the Hubble and HOD parameters included.  Moreover,
it clearly reflects the expected $\Omega_m/M_1=constant$ degeneracy
intrinsic to the halo mass function, though it is slightly
modified due the weak sensitivity of the survey volume to
$\Omega_m$.  The second eigenvector does not have a simple
interpretation \citep[though see Appendix in][]{rozoetal04}.\footnote{Curiously, 
we note that the top normal mode does not contain the degeneracy
direction $\alpha^2\sigma_8$ alluded to earlier.  Rather, both of the top two
eigenmodes have considerable $\alpha$ and $\sigma_8$ dependence, so
the $\alpha^2\sigma_8$ degeneracy is only recovered after marginalizing
over all other parameters.}
Hereafter, we refer to the top normal model in parameter space as the
generalized cluster normalization condition.

\begin{figure}[t]
\epsscale{1.2}
\plotone{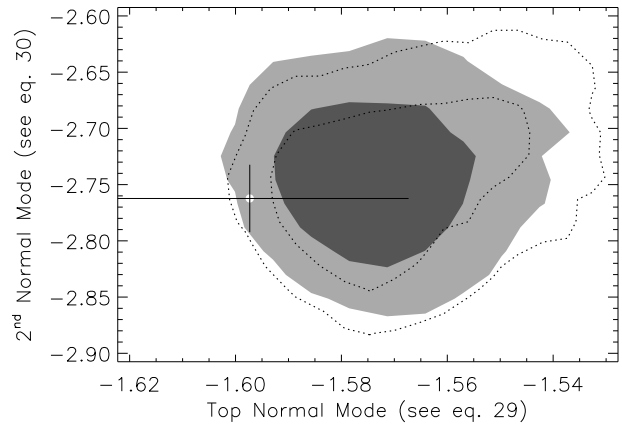}
\caption[Confidence Regions of the Top Two Normal Modes in Parameter Space
from Simulations]{Filled contours are the $68\%$ and $95\%$ confidence regions in the 
$\alpha-\sigma_8$ plane for the the two best-constrained parameter combinations.
The circle with
error bars marks the input simulation parameters, and the error bars are a $3\%$
uncertainty from the the HOD fits to the simulation. Again,
we find that our likelihood model successfully recovered the simulation
parameters.  Note that this is a very stringent test: the $1-\sigma$ 
error bars for these two top normal modes are $1\%$ and $5\%$ respectively. 
Also shown
above with dotted curves are the $68\%$ and $95\%$ confidence regions 
obtained when using WMAP and supernova like Gaussian priors $\Delta\Omega_h^2=0.01$
and $\Delta h=0.05$.  The apparent increase in the confidence regions is due to a slight
rotation of the likelihood function in parameter space due to the introduction these
cosmological priors.}
\label{fig:ssim_top2}
\end{figure}

We show the $68\%$ and $95\%$ confidence regions of these two
parameters for Mock A in Figure \ref{fig:ssim_top2}.  We
find that not only do we indeed recover the correct simulation
parameters, but that the associated statistical uncertainty is
extremely small, of order $1\%$ and $4\%$ for the top two normal modes
for our assumed survey of $1/8$ of the sky and redshift ranges
$z\in[0.13,0.25]$.  Given the small size of our error bars, the
excellent agreement between our statistical analysis and the true
simulation parameters is highly non-trivial.  In particular, it
explicitly demonstrates that \it if the selection function for the
maxBCG catalog can be tightly constrained, optically-selected cluster
samples can provide percent-level determinations of specific
combinations of cosmological and HOD parameters.\rm


It is also worth investigating to what extent our constraints can be
improved upon through the use of other cosmological probes.  In
particular, the CMB places strong constraints on $\Omega_m h^2$
\citep[see e.g.][]{huetal96,hudodelson02,dodelsonbook}, while
supernovae data puts strong constraints on the value of the Hubble
parameter $h$ \citep[see
e.g.][]{hstkeyproject}.  The reason these
two particular priors are interesting is that their values have
minimal or no dependence on the dynamical nature of dark energy.  That
is, these constraints do not depend on whether dark energy is a
cosmological constant or not.  Consequently, employing these priors
still allows us to use cluster abundances for studying the dark
energy.
Note that this is not the case for all priors.  For instance, the CMB
data can also provide priors on $\sigma_8$, provided the power
spectrum at last scattering is extrapolated to the present epoch using
a $\LCDM$ cosmology.  Clearly, such a prior is useless if one is
interested in constraining the behavior of the dark energy.  Indeed,
this is precisely why estimating $\sigma_8$ is an interesting problem:
deviations from the CMB interpolated value for $\sigma_8$ could signal
a failure of the $\Lambda$CDM model.

To investigate the impact that CMB and supernova like priors can have
on our results, we repeat the above analysis, but including now
Gaussian priors of width $\Delta\Omega_m h^2=0.01$ and $\Delta h=0.05$
centered on the simulation cosmology.  The width of these priors is
set by the current uncertainty in each of the two cosmological
parameters as constrained by the CMB and supernovae respectively
\citep[][]{wmap06}.  We find that including these cosmological priors
has minimal impact on how well our normal modes are constrained.  This
is shown in Figure \ref{fig:ssim_s8_aN}, where we find that the
$\alpha_N-\sigma_8$ degeneracy is only marginally reduced.  Even more
telling is Figure \ref{fig:ssim_top2}, where we show the confidence
regions of the normal modes found in the no priors case.  Note that
the confidence regions slightly increase rather than decrease due to
the rotation of the likelihood function in parameter space due to the
introduction of the priors.  Indeed, we find that including priors in
our analysis results in not just two but three highly constrained
modes at roughly $1\%$, $2\%$, and $5\%$ accuracy.  The first and the
third are almost identical to the normal modes found in the no priors
case, and the ones shown in Figure \ref{fig:ssim_top2}.  The second
normal mode, on the other hand, is largely parallel to the direction
of our priors.

In summary, we have found that local cluster abundances estimated from
large surveys can provide percent-level constraints on combinations of
cosmological and HOD parameters if the selection function,
i.e. completeness, purity, and the signal matrix, is known precisely.
Individual parameters cannot be constrained due to intrinsic
degeneracies in the data.  Finally, adding cosmological priors from
CMB and/or supernovae has minimal impact on the best-constrained
parameter combinations.


\subsection{Marginalizing Parameter Constraints Over Uncertainties in the Selection Function}
\label{sec:float_nuisance}

Consider now marginalization over uncertainties in the selection
function.  We found that the completeness, purity, and photometric
redshift error parameters were well constrained from the simulations,
and varying them through the range of values measured from the simulations 
had minimal impact on the estimated number counts.  Indeed, upon adopting top hat
priors corresponding to the $95\%$ regions of these parameters we find
that our results are largely identical to the ones presented in
\S \ref{sec:fixed_nuisance}.  Consequently, henceforth every
result we present is marginalized over the completeness, purity, and
photo-$z$ parameters.

The signal matrix parameters, on the other hand, are a different
story.  We saw in \S \ref{sec:signalmatrix} that the slope
$\beta$ of the mean relation between $\Nt$ and $\Nobs$ was relatively
well constrained, and that a Gaussian prior on $\beta$ of the form
$\beta=1.18\pm 5\%$ appears reasonable based on this set of
simulations.  Consequently, unless specially stated otherwise, we
shall assume this prior in all of our analysis.  We also saw, however,
that the amplitudes $B_0$ and $B_1$ of the signal matrix had large
systematic errors.  In fact, these errors are large enough that attempts
to constrain cosmology and HOD in the simulations using priors that covered
the whole range of selection functions observed in the simulations proved
unsuccessful.  We
have thus chosen to investigate how more moderate uncertainties in the
amplitudes affect our results, with an eye towards future work which
may improve our understanding of the cluster selection function.  To
do so, we ran MCMCs with $5\%$, $10\%$, $15\%$, and $20\%$ priors on
the signal matrix parameters $B_0$ and $B_1$ using the observed number
counts in Mock A.  Due to the computational effort involved in running
MCMCs for each model we consider, we focus on Mock A only.  There is
no particular reason why this realization was chosen over the other
two, and, based on our results from the previous section, we have no
reason to suspect that any one realization would lead to substantially
different results than the other two.  \footnote{Marginalizing over
  the signal matrix parameters in our analysis also gave rise to
  numerical difficulties in the realization of the MCMC.  A
  description of these problems and how they were overcome is given in
  Appendix \ref{sec:implementation}.}

\begin{figure}[t]
\epsscale{1.2}
\plotone{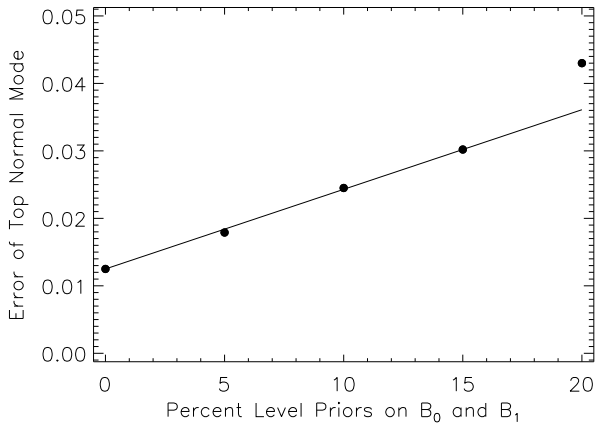}
\caption[Normal Mode Errors as a Function of Uncertainty in the Cluster Selection
Function]{Sensitivity of the error of the best-constrained normal mode in
parameter space to uncertainties in the cluster selection function.  This
mode corresponds to a generalized cluster normalization condition, and
both its amplitude and direction are fairly robust for up to $\approx 15\%$ 
uncertainties in
the cluster selection function, characterized in this case by the amplitudes
$B_0$ and $B_1$ (see \S \ref{sec:signalmatrix}).  Filled circles
assume a $5\%$ prior on the slope $\beta$, while the triangle is obtained
assuming a $10\%$ prior on $\beta$.  Finally, the square marks the error
on the generalized cluster normalization condition when including CMB and
supernova like priors on $\Omega_m h^2$ and $h$ (compare to Figure
\ref{fig:ssim_s8_aN}.  The slight increase in the uncertainty is due to the
change in orientation of the likelihood function.}
\label{fig:top_mode_errors}
\end{figure}

Because of the large degeneracies in parameter space, we have chosen
to focus on the two best-constrained modes in parameter space to
quantify the sensitivity of our results to uncertainties in the values
of $B_0$ and $B_1$.  We find that the best-constrained normal mode is
robust to $\approx 15\%$ uncertainties in the selection function. In
particular, the direction of the mode remains constant, and the
uncertainty in the parameter increases linearly with the width of the
assumed priors, as illustrated in Figure \ref{fig:top_mode_errors}.
By $20\%$ uncertainties in the amplitudes $B_0$ and $B_1$, however,
the top mode has rotated away slightly, and its uncertainty starts
growing faster than linearly with the width of the amplitude priors.
Nevertheless, it is remarkable that even with uncertainties as large
as $20\%$ in the cluster selection function we can recover the top
normal mode in parameter space to better than $5\%$ accuracy.


To investigate how sensitive our results were to our assumptions about
$\beta$, we also considered the case in which all signal matrix
parameters (including the slope $\beta$) were known to $10\%$
accuracy.  The corresponding error on the top normal mode for this
case is shown in Figure \ref{fig:top_mode_errors} as a triangle, and
demonstrates that there is little loss of information by the
additional uncertainty in $\beta$.  It is worth noting that the two
model parameters that are most closely aligned with the top normal
mode are $\alpha$ and $\sigma_8$.  Consequently, constraints in the
$\alpha-\sigma_8$ are relatively robust to uncertainties in the signal
matrix, as shown in Figure \ref{fig:ssim_s8_aN}.

We now turn our attention to the behavior of the second
best-constrained mode, which we find is \it not \rm stable to
uncertainties in the signal matrix.  Specifically, we find that there
is a factor of two increase in the error of this mode in going from
fixed nuisance parameters to $5\%$ uncertainties in the signal matrix.
Moreover, the \it direction \rm of the second best-constrained normal
mode is substantially different between the two cases.  Curiously, as
we increased the width of our prior on the amplitudes $B_0$ and $B_1$,
we found that the second best-constrained mode remained relatively
constant both in direction and width, suggesting the large difference
between the fixed nuisance parameter case and the $5\%$ priors case
was driven largely by the uncertainties in the slope $\beta$.  We
tested this scenario by running an additional chain with $10\%$ priors
on all signal matrix parameters, and found that, indeed, with the new
priors for $\beta$ the second best-constrained mode was severely
affected, both in terms of the direction and the percent-level
accuracy with which it could be recovered.  We conclude that there is
a large degeneracy between cosmological and HOD parameters and the
slope $\beta$.  This is an important, though rather unfortunate,
result, as it implies that to fully recover the constraining power of
large local cluster samples, the slope of the relation between $\Nobs$
and $\Nt$ must be known to high accuracy.

In light of these results, it is worth returning to the question of
whether or not CMB and supernova priors on cosmological parameters
could substantially alter our conclusions.  To test this, we ran an
addition MCMC using CMB and supernova like Gaussian priors
$\Delta\Omega_mh^2=0.01$ and $\Delta h=0.05$, assuming $15\%$
uncertainties in $B_0$ and $B_1$, and our default $5\%$ level
uncertainty in $\beta$.  We mark the corresponding uncertainty on the
cluster normalization condition in Figure \ref{fig:top_mode_errors} as
a square.  Note that the error on the cluster normalization
condition slightly increases upon inclusion of the prior.  This is
again indicative of an overall distortion of the orientation of the
likelihood surface in parameter space upon inclusion of the priors.
Indeed, upon including the priors, we found that the best-constrained
mode was no longer the cluster normalization condition, but rather
falls close to the direction of the assumed priors.  The generalized
cluster normalization condition then becomes the second
best-constrained eigenmode, and was itself slightly rotated relative
to the fixed nuisance parameters case.  The next best-constrained
eigenmode was found to be unstable to the introduction of cosmological
priors.


\section{Summary and Discussion}
\label{sec:conclusions}

In this work we have introduced a general framework for characterizing
the selection function of optical cluster finding algorithms.  The
fundamental assumption in our method is that the scatter in the
mass-observable relation for a cluster finding algorithm can be split
into an intrinsic scatter, and an observable scatter due to the
imperfection of the cluster finding algorithm.  We show that the
inability to fully characterize catastrophic errors in richness
assignments naturally gives rise to the concepts of purity and
completeness in quantitative form.  These definitions of purity and
completeness are well defined and are particularly well suited to
cosmological abundance analyses.

This method could potentially be applied to characterizing the
selection function and to cosmological parameter estimation for a wide
range of current and future cluster samples.  Here, we have
demonstrated its utility by application to the maxBCG cluster finding
algorithm \citep[][]{koesteretal06a}, run on mock galaxy catalogs
produced using three different realizations of the ADDGALS
prescription for connecting a realistic galaxy population to large
dissipationless simulations, of comparable volume to the SDSS data
sample (detailed in \citealt[][]{wechsleretal07}).  In a companion
paper \citep{rozoetal07b}, we apply this method to the SDSS maxBCG
cluster catalog \citep{koesteretal06b}.

By matching the input halos to the detected clusters, we have
quantitatively calibrated the maxBCG selection function in each of the
three mock catalogs, and demonstrated that with knowledge of this
selection function we can accurately recover the underlying cosmology
and HOD parameters of the simulations to within the intrinsic
degeneracies of the data.  Moreover, we have shown that this is still
the case when the selection function is only known to $\approx 15\%$
accuracy, though the uncertainty in the recovered parameters starts
growing quickly after that.

We conclude that it is possible to provide tight cosmological and HOD
constraints using optically-selected cluster catalogs, but doing so
requires a better constrained cluster selection function that we
currently have. This is an important and non-trivial result: it
explicitly shows that the popular view that projection effects present
an insurmountable obstacle for precision cosmology with with
optically-selected cluster catalogs is no longer the case.  We have
demonstrated that the maxBCG cluster catalog is highly complete and
pure, and, more importantly, that any such effects can be incorporated
into our cosmological parameter analysis through a detailed
calibration of the cluster selection function.  Provided the selection
function is known with relative accuracy, optical cluster catalogs can
be useful tools for precision cosmology.

In the present work, we have not made an exhaustive attempt to
characterize the uncertainties in the cluster selection function.
Here, we investigated three realizations of the empirically-motivated
galaxy biasing scheme ADDGALS, applied to one cosmological model, the
large, low resolution Hubble Volume simulation.  Although all three
realizations provide a reasonable representation of galaxies in the
local Universe, including realistic luminosity and color evolution and
clustering properties, and a red sequence population that is a good
match to maxBCG, the three simulations had different HOD descriptions.
Our results suggest that the cluster selection function depends to
some extent on the specific HOD of the simulation, at least for the
richness measurements we considered.  To mitigate this uncertainty in
our analysis of the maxBCG data, in \citet[][]{rozoetal07b}, we
perform the analysis assuming only that the shape of the selection
function is the same in both the simulations and the data, and greatly
relaxing the prior on the slope $\beta$ relating the mean observed and
intrinsic richness.


We end now by considering the obvious question: can this situation
improve?  We remain optimistic that future work will allow tighter and
more robust constraints on the maxBCG selection function than are
presented here, which will allow us to maximize the power of the large
maxBCG data set.  We are proceeding along three fronts:
\begin{itemize}
\item Improved richness definitions. A robust calibration for
  $P(\Nobs|\Nt)$ \it for arbitrary definitions of $\Nt$ and $\Nobs$
  \rm may indeed be hard to come by, it is entirely plausible that we
  can refine our definitions of halo and cluster richness to
  considerably improve our understanding of the selection function.
  For instance, in this work, no attempt was made to make $\Nobs$ an
  unbiased estimator of $\Nt$, so something as simple as including a
  color cut in our definition of $\Nt$ could significantly improve our
  model.  If one can define $\Nobs$ such that,
  by construction, $\avg{\Nobs|\Nt}=\Nt$, not only will the number of
  nuisance parameters immediately go down by two, but also some of the
  large degeneracies we uncovered in this work will become irrelevant.
\item A detailed characterization of the variance between a range of
  models.  A crucial question is whether the selection function
  calibration is robust to changes not only in the halo occupation of
  the galaxies but also in the cosmological parameters of the
  underlying simulation.  Although we didn't explore this directly in
  the mocks investigated here, we were generous in the range of galaxy
  populations applied to the simulations.  Scatter between selection
  function parameters will likely go down if we apply further
  observational constraints on the galaxy populations.  We then intend
  a wide exploration of parameter space after these constraints have
  been applied.
\item The addition of mass calibration data from maxBCG itself.
  Information on the mass scale is directly available from both
  stacked lensing measurements (\citealt{sheldonetal07}, Johnston et
  al, in preparation), stacked X-ray measurements \citep[][]{rykoffetal07}, 
  and from the velocity dispersions of the
  galaxies in clusters \citep[][]{beckeretal07}.  These data
  provide substantial additional constraints on combinations of our
  selection function parameters, and will allow us to use weaker
  priors in both the selection function and cosmological parameter space.
\end{itemize}

\acknowledgments ER would like to thank Scott Dodelson and Andrey
Kravtsov for a careful reading of an earlier version of the
manuscript, and for many illuminating comments that have greatly
improved the quality and presentation of this work.  ER would also
like to thank Wayne Hu, Zhaoming Ma, Andrew Zentner, and Marcos Lima
for useful conversations.  This work was carried out as part of the
requirements for graduation at The University of Chicago.  ER was partly
supported the Center for Cosmology and Astro-Particle Physics (CCAPP)
at The Ohio State University.  ER was also funded in part by the Kavli
Institute for Cosmological Physics (KICP) at The University of Chicago.  RHW was
primarily supported by NASA through a Hubble Fellowship awarded by the
Space Telescope Science Institute, which is operated by the
Association of Universities for Research in Astronomy, Inc, for NASA,
under contract NAS 5-26555.  RHW was also supported in part by the 
U.S. Department of Energy under
contract number DE-AC02-76SF00515.
AEE was supported in part by NASA grant NAG5-13378, by NSF
ITR grant ACI-0121671, and by the Miller Foundation for Basic
Research in Science at UC, Berkeley. 
T. McKay, A. Evrard, and B. Koester gratefully acknowledge support from 
NSF grant AST 044327.
This study has used data from the Sloan
Digital Sky Survey (SDSS, http://www.sdss.org/).
Funding for the SDSS and SDSS-II has been provided by the
Alfred P. Sloan Foundation, the Participating Institutions, the National Science
Foundation, the U.S. Department of Energy, the
National Aeronautics and Space Administration, the Japanese
Monbukagakusho, the Max Planck Society, and the Higher Education Funding 
Council for England.  
Some of the simulations in
this paper were realized by the Virgo Supercomputing
Consortium at the Computing Centre of the Max-Planck
Society in Garching and at the Edinburgh Parallel Computing
Centre. Data are publicly available at
www.mpa-garching.mpg.de/NumCos. 
This work made extensive
use of the NASA Astrophysics Data System and of the {\tt astro-ph}
preprint archive at {\tt arXiv.org}.

\bibliographystyle{apj}
\bibliography{v8a}

\newcommand\AAA[3]{{A\& A} {\bf #1}, #2 (#3)}
\newcommand\PhysRep[3]{{Physics Reports} {\bf #1}, #2 (#3)}
\newcommand\ApJ[3]{ {ApJ} {\bf #1}, #2 (#3) }
\newcommand\PhysRevD[3]{ {Phys. Rev. D} {\bf #1}, #2 (#3) }
\newcommand\PhysRevLet[3]{ {Physics Review Letters} {\bf #1}, #2 (#3) }
\newcommand\MNRAS[3]{{MNRAS} {\bf #1}, #2 (#3)}
\newcommand\PhysLet[3]{{Physics Letters} {\bf B#1}, #2 (#3)}
\newcommand\AJ[3]{ {AJ} {\bf #1}, #2 (#3) }
\newcommand\aph{astro-ph/}
\newcommand\AREVAA[3]{{Ann. Rev. A.\& A.} {\bf #1}, #2 (#3)}

\appendix

\section{Summary of Equations}
\label{app:summary}

For reference, we summarize below all the formulae that quantitatively describe
our model.  Let $\bm{N}=\{N_1,...,N_M\}$ be the number of clusters in richness
bins $1$ through $M$, and $\bm{p}$ be the parameters characterizing cosmology,
HOD, purity, completeness, and the signal matrix $\ps$.
The likelihood function $\lk(\bm{p}|\bm{N})$ is given by
\begin{equation}
\lk(\bm{p}|\bm{N}) \propto \frac{1}{\sqrt{(2\pi)^M \mbox{det}(C)}}
	\exp\left\{ - \frac{1}{2}(\bm{N} - \avg{\bm{N}}) \cdot C^{-1}
	\cdot (\bm{N}-\avg{\bm{N}})\right\}.
\end{equation}
Above, $\avg{\bm{N}}$ is the expectation value for the data vector, and $C$ is 
the correlation matrix of the observables.  The number of clusters in a given
richness bin $a$ is given by
\begin{equation}
\avg{\Na} = \int d\zh d^2\bn_h\ \chi^2\frac{d\chi}{d\zh}\avg{\na}\avg{\varphi|\zh}\avg{\Theta|\bn_h}
\end{equation}
where $\chi$ is the comoving distance to redshift $\zh$.  The function $\avg{\varphi|\zh}$
is the effective redshift selection function of the survey, and is given by
\begin{equation}
\avg{\varphi|\zh}= \int d\zc\ \rho(\zc|\zh)\varphi(\zc)
\end{equation}
where $\varphi(\zc)=1$ if the photometric cluster redshift $\zc\in[\zmin,\zmax]$ 
and $\varphi(\zc)=0$ otherwise, and $\zmin$ and $\zmax$ define the redshift selection
criteria for the survey.  Likewise, the function $\avg{\Theta|\bn_h}$ is the effective 
angular mask of the survey, and is given by
\begin{equation}
\avg{\Theta|\bn_h} \int d^2\bn_c \Theta(\bn_c)\rho(\bn_c|\bn_h)
\end{equation}
where $\Theta(\bn_c)$ is the angular mask of the survey.  In this work, we
have assumed $\rho(\bn_c|\bn_h)=\delta(\bn_c-\bn_h)$.
Finally, $\avg{\na}$ is the expected comoving
cluster number density as a function of redshift, and is given by
\begin{equation}
\avg{n_a} = \int dm \nm \avg{c\psia/p|m}.
\end{equation}
where $\avg{dn/dm}$ is the halo mass function, and 
the quantity $\avg{c\psia/p|m}$ represents the mass binning of the survey, given
by
\begin{equation}
\avg{c\psia/p|m} = \sum_{\Nt} c(\Nt)\tilde\psia(\Nt)P(\Nt|m).
\end{equation}
where $c(\Nt)$ is the completeness function, 
and $P(\Nt|m)$ is the HOD.  $\tilde\psia$ is the binning function in terms
of $\Nt$, which is related to the top hat richness binning function 
$\psia(\Nobs)$ via
\begin{equation}
\tilde\psia(\Nt) = \sum_{\Nobs} \frac{\psia(\Nobs)}{p(\Nobs)}P_s(\Nobs|\Nt).
\end{equation}
where the sum is over all $\Nobs$ values within the signal band, $p(\Nobs)$
is the purity function, and $P_s(\Nobs|\Nt)$ is the signal matrix.

We also summarize below the 
equations describing the correlation matrix $C$.  
In particular, the Poisson contribution to the correlation
matrix takes on the form
\begin{equation}
(\corr)_P = \delta_{a,a'}\int(\chi^2 \frac{d\chi}{d\zh}d\zh d\bn_h) 
	\int dm \nm \avg{c\psia/p^2|m} \avg{\Theta|\bn_h}\avg{\varphi|\zh}.
\end{equation}
Note that, schematically,
this contribution takes on the form $\corr \approx \delta_{a,a'}\avg{\Na}/p$.
This is exactly what we would expect: if $N$ is the number of observed
clusters and $\Ns$ the number of signal clusters, then $N=\Ns/p$, so
$\mbox{Var}(N)=\mbox{Var}(\Ns)/p^2=N/p$.  
To this Poisson contribution we must also add the sample variance term
\begin{equation}
(\corr)_S = \avg{bG\Na}\avg{bG\Nap} \sigma^2(V)
\end{equation}
where 
\begin{equation}
\avg{bG\Na} = V\left[\int dm\ b(m)\nm\avg{c\psia/p|m}G \right]_{\bar z},
\end{equation}
$V$ is the survey volume,
and $\sigma^2(V)$ is the rms variance of the linear density field over the sample volume
probed.  An additional contribution to the correlation matrix arises from the
stochasticity of the mass-richness relation, which leads to uncertainties in the
precise mass binning of the cluster sample.  This contribution takes on the form
\begin{eqnarray}
(\corr)_B & = &\int d\zh d^2\bn_h\ \chi^2\frac{d\chi}{d\zh}\avg{\varphi|\zh}\avg{\Theta|\bn_h} 
	\avg{\varphi|\zh} \\
& & \times \int dm\ \nm \left[
	\delta_{a,a'}\avg{c\psia/p^2|m} - \avg{c\psia/p|m}\avg{c\psiap/p|m} \right].
\end{eqnarray}
Note that in the above expression it is evident that neighboring bins are always
negatively correlated, reflecting the fact that halos scattering into a given
bin $a$ must have scattered out of some other bin $a'$.
Also, there is an additional contribution to do photometric redshift uncertainties,
which reduces to
\begin{equation}
(\corr)_Z = \delta_{a,a'} \int d\zh d^2\bn_h\ \chi^2\frac{d\chi}{d\zh}\avg{\Theta|\bn_h}
	\left[ \avg{\varphi|\zh}-\avg{\varphi|\zh}^2 \right]
		\int dm\ \nm \avg{c\psia/p^2|m}.
\end{equation}
Finally, there are uncertainties due to the stochastic nature of the purity function,
which take on the form
\begin{equation}
(\corr)_p = \delta_{a,a'} \int d\zh \frac{dV}{d\zh} \avg{\varphi|z}
	\int dm \nm \avg{c\psia(1-p)/p|m}.
\end{equation}
Note that when the purity is exactly equal to one, the purity contribution to
the correlation matrix vanishes, as it should.


\section{Matching Algorithms Considered in this Work}
\label{app:matchmodels}

The type of membership matching algorithms we considered can be broadly grouped
into one of two categories: deterministic matching algorithms 
and probabilistic matching algorithms.  We considered four deterministic matching
algorithms:

\begin{enumerate}
\item \it Maximum shared membership matching: \rm Each halo is matched
  to the cluster containing the largest number of galaxies belonging
  to the halo.
\item \it BCG matching: \rm A halo is matched to the cluster
  containing the halo's BCG.
\item \it Exclusive maximum shared membership matching: \rm Halos are
  first rank ordered according to their richness $\Nt$.  Starting with
  the richest halo, its cluster match is found through maximum shared
  membership matching.  The matched cluster is then removed from the
  list of candidate matches for all other halos, and the procedure is
  iterated.
\item \it Exclusive BCG matching: \rm This is exactly analogous to
  exclusive maximum shared membership, only BCG matching is used to
  match halos to clusters at each step.
\end{enumerate}

In addition to these four matchings, we considered what we have called
probabilistic matching schemes, which are essentially generalizations
of the deterministic matching schemes. The idea is as follows: imagine
listing all halos and all clusters in a two column format, so that the
left column contains all halos and the right column contains all
clusters.  Pick a cluster $\alpha$, and draw a line connecting halo
$\alpha$ to all clusters $\beta$ which share members with halo
$\alpha$.  Let then $f_{\alpha\beta}$ be the fraction of galaxies in
halo $\alpha$ contained in cluster $\beta$.  Maximum shared membership
matching consists of selecting the cluster $\beta$ which maximizes
$f_{\alpha\beta}$ with probability one.  More generally, however, one
could imagine replacing this probability with some other probability
function $p(f_{\alpha\beta})$, an obvious choice being
$p(f_{\alpha\beta})=f_{\alpha\beta}$ which we refer to as \it
proportional random matching\rm .  Another possible choice is \it
random membership matching, \rm where we set
$p(f_{\alpha\beta})=1/N_{\alpha}$ where $N_{\alpha}$ if the total
number of candidate cluster matches for halo $\alpha$.

Consider now the probability matrix $\pnn$ in the case of probabilistic matching.
One has that
\begin{equation}
\pnn = \sum_\alpha P(\Nobs|\alpha)P(\alpha|\Nt)
\end{equation}
where $P(\Nobs|\alpha)$ is the probability that halo $\alpha$ be matched
to a cluster with $\Nobs$ galaxies, and $P(\alpha|\Nt)=\delta_{N_{\alpha,\Nt}}/N(\Nt)$ 
is the probability of picking
halo $\alpha$ at random from a the set of $N(\Nt)$ halos of richness $\Nt$.  All that
remains is computing the probability $P(\Nobs|\alpha)$, which is simply given by
\begin{equation}
P(\Nobs|\alpha)=\sum_\beta \delta_{R_\beta,\Nobs} p(f_{\alpha\beta}).
\end{equation}
Putting it altogether we find
\begin{equation}
P(\Nobs|\Nt) = \frac{1}{N(\Nt)} \sum_{\alpha,\beta} 
	\delta_{R_\alpha,\Nt}\delta_{R_\beta,\Nobs}p(f_{\alpha\beta}).
\label{eq:prob}
\end{equation}

Figure \ref{fig:matching} plots the cost function $\Delta(\Nobs)$ for
each of the two exclusive matching algorithms, and for the
non-exclusive maximum shared membership matching in the case of the
\it s \rm simulation.  The corresponding plots for the other
simulations are quite similar.  The non-exclusive BCG matching and the
probabilistic matchings give results almost identical to the
non-exclusive maximum shared membership matching case, and are very
much worse than any of the exclusive matching algorithms,
demonstrating that enforcing a one to one matching between halos and
clusters is of paramount importance.  It is worth noting that maxBCG
does not enforce exclusive galaxy membership for cluster, that is, a
galaxy can be a member of more than one cluster.

We focus now on the exclusive matching algorithms.  In particular,
each of these algorithms has an upturn in the cost function at high
richness.  This upturn is due to what we called noise in \S
\ref{sec:model}, i.e. catastrophic errors in the cluster-finding
algorithm.  These catastrophic errors are in reality quite rare, so
the mixing induced by non-zero matrix elements due to catastrophic
errors generate an unrealistically large amount of mixing between low
richness halos and high richness clusters.  Consequently, we believe
the upturn at high richness is unphysical.  Given this assumption,
based on Figure \ref{fig:matching} we chose exclusive maximum shared
membership matching as our fiducial halo-cluster matching algorithm
for the remainder of this work.\footnote{The fact that $\Delta$ is
  lower at high richness for exclusive BCG matching suggests that one
  try a hybrid matching algorithm for which exclusive BCG matching is
  used at high richness, and exclusive maximum shared membership
  matching is used at low richness.  The transition richness $\Nt$
  between these two matchings would then be a tunable parameter.  We
  found that requiring this hybrid algorithm to perform comparably to
  the exclusive maximum shared membership algorithm in the low
  richness regime resulted in a transition richness $\Nt$ high enough
  that the hybrid algorithm became effectively identical to the
  exclusive maximum shared membership algorithm.}


\section{MCMC Implementation}
\label{sec:implementation}

Evaluation of the likelihood function in a multi-dimensional parameter
space can be extremely time consuming.  Fortunately, not all of
parameter space needs to be sampled, only the high likelihood regions.
We achieve this through the use of Monte Carlo Markov Chains (MCMC),
drawing heavily on the work by \citet[][]{dunkleyetal05} \citep[for
some introductory material on MCMC methods see e.g.][]{wasserman04}.

In this work, we require not only to find the point of maximum
likelihood, but also the $68\%$ and $95\%$ probability likelihood
contours.  \citet[][]{dunkleyetal05} showed that the number of points
$N$ in a chain necessary to recover a $p$ confidence region with
accuracy $q$ is $N\approx 3.3D/q^2/(1-p)$ where $D$ is the
dimensionality of the parameter space.  When we hold our nuisance
parameters fixed to the simulation calibrated values, the total number
of free parameters is five, so the number of points in the chain
necessary to recover the $95\%$ confidence regions with $10\%$
accuracy is $N\approx 3\cdot 10^4$.  We choose $N=5\cdot 10^4$ steps
as the default length of our chains in this case.  When fitting for
all 13 parameters (cosmology+HOD+nuisance), the expected minimum
number of points required to appropriately sample the likelihood
function is thus $N\approx 10^5$, which we adopt as our default length
for the chains.

Now, in order to achieve convergence quickly, some take must be taken
to ensure that the chain is close to optimal.  In particular, both the
direction and size of the steps in the MCMC need to be carefully
chosen.  Here, we follow \citet[][]{dunkleyetal05}, and take random
steps along the principal components of the parameter correlation
matrix.  For each normal mode, we chose a step size $2\sigma/\sqrt{D}$
where $\sigma$ is the eigenvalue of the principal component and $D$ is
the number of parameters (either $4$ or $\approx 10$ in our case).
This step size is slightly smaller than the value reported in
\citet[][]{dunkleyetal05}, and is chosen because, as shown by
\citet[][]{dunkleyetal05}, for problems with a large number of
parameters, it is in general better to err on the low side when
estimating the optimal step size.

Finally, in order to be able to take steps along the principal
components of the correlation matrix, we must first estimate the
correlation matrix.   To do so, we follow an iterative procedure where 
we start our MCMC in a region of
relatively high likelihood, and use the first $1000$ steps to estimate
the correlation matrix varying only cosmological and HOD parameters to
resolve any intrinsic degeneracies of the model.  For the next $2000$
steps we introduce the parameters characterizing the signal matrix as
additional degrees of freedom, and estimate the new correlation
matrix.  We then allow the completeness, purity, and photo-$z$'s
parameters to vary, and use $3000$ steps to estimate the correlation
matrix.  We make one final iteration of $4000$ steps and re-estimate
the correlations matrix, which is used to take an additional $10^5$
steps which constitute our chain for the purposes of parameter
estimation.  When keeping nuisance parameters fixed, we use only two
training runs to estimate the correlation matrix, the first with
$1000$ steps, and then an iteration with $2000$ steps.  We checked all
our chains for convergence with the tests described in
\citet[][]{dunkleyetal05}.

We found that letting the signal matrix parameters float in our MCMCs
introduced a very serious numerical difficulty.  In particular, we
found that in our model there was a small region of the $13$
dimensional space where the estimated correlation matrix of the
cluster number counts becomes singular, causing the likelihood
function to blow up.  We found this difficulty arose due to detailed
numerical cancellation between the Poisson and binning contributions
to the correlation matrix.  Consequently, we decided to approximate
the correlation matrix by replacing the Poisson contribution estimated
from our model by $1/\Na$ where $\Na$ are the \it observed \rm number
counts.  In high likelihood regions, one has $\avg{\Na}\approx \Na$
and hence the approximation should be valid.  Indeed, we explicitly
checked that we could reproduce all of our results when keeping the
signal matrix parameters fixed while employing this approximation.

\begin{figure}[t]
\epsscale{1.2}
\plotone{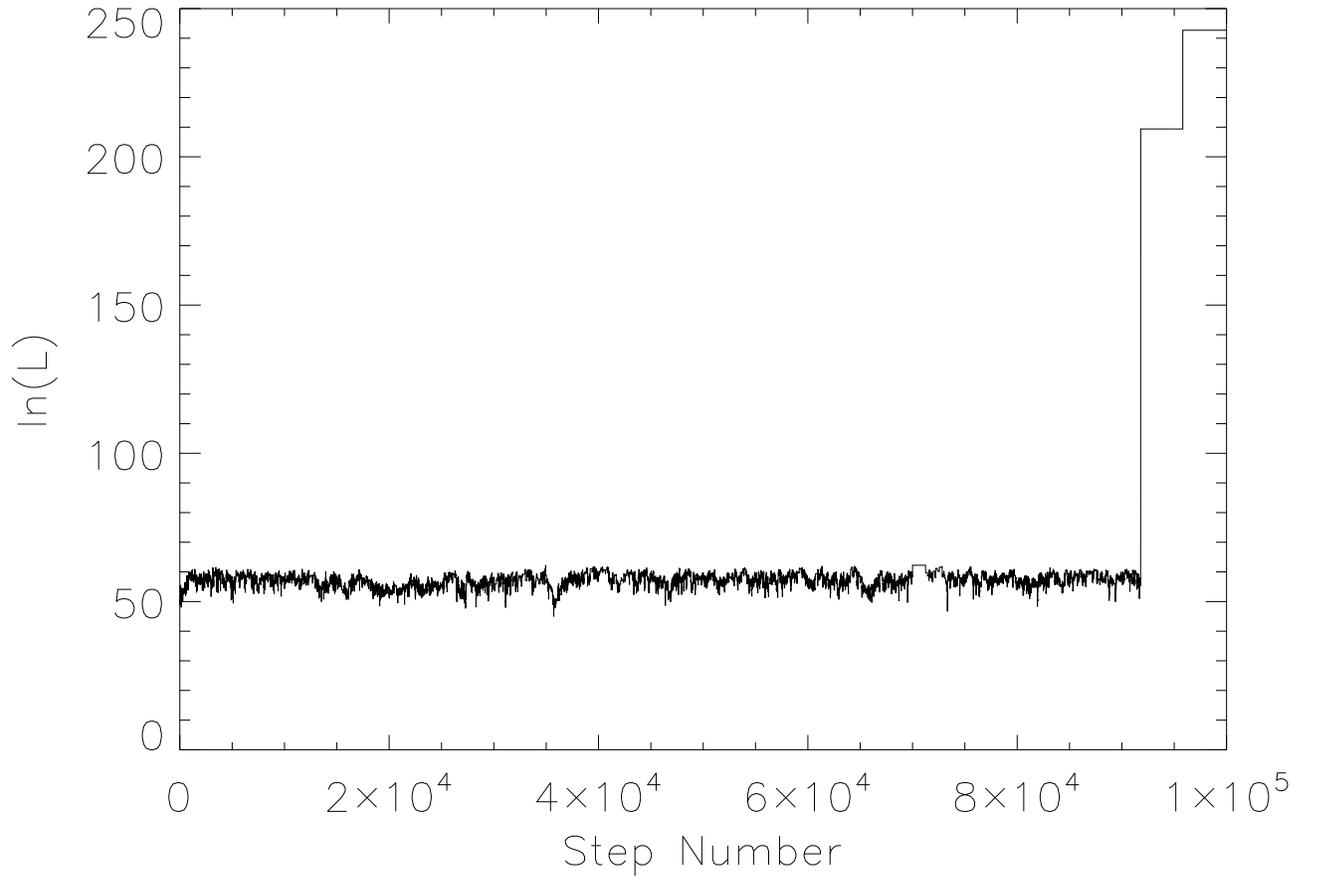}
\caption[Sample MCMC Output Illustrating Singularity in Parameter Space]{An 
MCMC output illustrating the singularity of the correlation matrix
in a small region of parameter space.  If the estimated correlation matrix of
the cluster number counts becomes singular, the likelihood function is infinite.
To avoid the corresponding small problem region in parameter space we introduce
a likelihood cut.  See text for more discussion.}
\label{fig:blowup}
\end{figure}

The end result of our approximation was to greatly reduce the volume
of parameter space where the cluster counts correlation matrix became
singular.  A typical output from a chain computed with the above
likelihood is shown in Figure \ref{fig:blowup}.  The fact the singular
region in parameter space is small can be seen from the fact that the
chain ran for $\approx 9\cdot 10^4$ steps before encountering the
singularity.  Moreover, once the chain stepped into this region, it
froze, demonstrating that virtually every step took it outside the
singularity.  To avoid this behavior, we can simply carve out the
singular region of parameter space, which we do by introducing a cut
on the likelihood.  In particular, we set $\lk=0$ for any region where
our model results in an unrealistically large likelihoods.  It is
worth noting that introducing such a cut without affecting the
performance of the chain is not difficult.  For instance, in Figure
\ref{fig:blowup}, we see that the likelihood ratio between the
singular region and the physical region is $\approx 10^{65}$.


\end{document}